\documentclass[12pt]{article}
\usepackage{graphicx}
\usepackage{amsmath}
\usepackage{amssymb}
\usepackage{xspace}

\usepackage{scicite}

\usepackage{times}

\newenvironment{sciabstract}{%
\begin{quote} \bf}
{\end{quote}}

\newcounter{lastnote}

\newcommand{\tess}{\textit{TESS}\xspace}

\newcommand{\exoplanet}{\textsf{exoplanet}\xspace}

\newcommand{\giants}{\textsf{giants}\xspace}

\newcommand{\hoststar}{TIC 365102760\xspace}
\newcommand{\starmass}{$1.21\pm0.06$ $M_\odot$\xspace}
\newcommand{\starradius}{$3.23\pm0.12$ $R_\odot$\xspace}
\newcommand{\teff}{$4694\pm100$ K\xspace}
\newcommand{\feonh}{$0.53\pm0.06$ dex\xspace}

\newcommand{\age}{$7.2\pm 1.4$ Gyr\xspace}
\newcommand{\starrho}{$0.041\pm0.006$ $\rho_\odot$\xspace}
\newcommand{\logg}{$3.51\pm0.03$ dex\xspace}

\newcommand{\planet}{TIC 365102760 b\xspace}
\newcommand{\planetmass}{$19.2\pm4.2$ $M_\oplus$\xspace}
\newcommand{\planetradius}{$6.21\pm0.76$ $R_\oplus$\xspace}
\newcommand{\period}{$4.21285367\pm0.00000074$ d\xspace}
\newcommand{\transittime}{$2458684.146\pm0.011$}

\title{An unlikely survivor: a low-density hot Neptune orbiting a red giant star}

\author
{Samuel Grunblatt,$^{1\ast,2}$ Nicholas Saunders,$^{3}$ , Daniel Huber$^{3,4}$, \\ 
Daniel Thorngren$^{1}$, Shreyas Vissapragada$^{5}$,  Stephanie Yoshida$^{5}$, \\ Kevin Schlaufman$^{1}$, Steven Giacalone$^{6}$, Mason MacDougall$^{7}$, Ashley Chontos$^{8}$, \\
Emma Turtelboom$^{6}$, Corey Beard$^{9}$, Joseph M. Akana Murphy$^{10}$, Malena Rice$^{11,12}$, \\ Howard Isaacson$^{6,13}$,
Ruth Angus$^{2,14,15}$, Andrew W. Howard$^{16}$ \\
\\
\normalsize{$^{1}$Department of Physics and Astronomy, Johns Hopkins University,} \\
\normalsize{3400 N Charles St, Baltimore, MD 21218, USA}\\
\normalsize{$^{2}$Department of Astrophysics, American Museum of Natural History,}\\
\normalsize{200 Central Park West, Manhattan, NY 10024, USA}\\
\normalsize{$^{3}$Institute for Astronomy, University of Hawaii at M\=anoa,}\\
\normalsize{2680 Woodlawn Drive, Honolulu, HI 96822, USA}\\
\normalsize{$^{4}$Sydney Institute for Astronomy (SIfA), School of Physics, University of Sydney, }\\
\normalsize{Camperdown, NSW 2006, Australia}\\
\normalsize{$^{5}$Center for
Astrophysics $|$ Harvard \& Smithsonian}\\
\normalsize{60 Garden Street, Cambridge, MA 02138, USA}\\
\normalsize{$^{6}$Department of Astronomy,}\\
\normalsize{University of California Berkeley, Berkeley, CA 94720, USA}\\
\normalsize{$^{7}$Department of Physics \& Astronomy, } \\
\normalsize{University of California Los Angeles, Los Angeles, CA 90095, USA}\\
\normalsize{$^{8}$Department of Astrophysical Sciences, Princeton University, } \\
\normalsize{4 Ivy Lane, Princeton, NJ 08544, USA}\\
\normalsize{$^{9}$Department of Physics \& Astronomy, } \\
\normalsize{University of California Irvine, Irvine, CA 92697, USA}\\
\normalsize{$^{10}$Department of Astronomy and Astrophysics, } \\
\normalsize{University of California, Santa Cruz, CA 95064, USA}\\
\normalsize{$^{11}$Department of Physics and Kavli Institute for Astrophysics and Space Research,}\\
\normalsize{Massachusetts Institute of Technology, Cambridge, MA 02139, USA}\\
\normalsize{$^{12}$Department of Astronomy, Yale University, New Haven, CT 06511, USA}\\
\normalsize{$^{13}$Centre for Astrophysics, University of Southern Queensland,}\\
\normalsize{ Toowoomba, QLD, Australia}\\
\normalsize{$^{14}$Center for Computational Astrophysics, Flatiron Institute,}\\
\normalsize{162 5$^\mathrm{th}$ Avenue, Manhattan, NY 10010, USA}\\
\normalsize{$^{15}$Department of Astronomy, Columbia University,}\\
\normalsize{550 West 120$^\mathrm{th}$ Street, New York, NY, USA}\\
\normalsize{$^{16}$Department of Astronomy, California Institute of Technology,}\\
\normalsize{Pasadena, CA 91125, USA}\\
\\
\normalsize{$^\ast$Corresponding author E-mail:  sgrunbl2@jhu.edu}
}

\date{}

\begin{document} 


\baselineskip24pt


\maketitle


\begin{sciabstract}

Hot Neptunes, gaseous planets smaller than Saturn ($\sim$ 3-8 R$_\oplus$) with orbital periods less than 10 days, are rare. Models predict this is due to high-energy stellar irradiation stripping planetary atmospheres over time, often leaving behind only rocky planetary cores. We present the discovery of a 6.2 R$_\oplus$(0.55 R$_\mathrm{J}$), 19.2 M$_\oplus$(0.060  M$_\mathrm{J}$) planet transiting a red giant star every 4.21285 days. The old age and high equilibrium temperature yet remarkably low density of this planet suggests that its gaseous envelope should have been stripped by high-energy stellar irradiation billions of years ago. The present day planet mass and radius suggest atmospheric stripping was slower than predicted. Unexpectedly low stellar activity and/or late-stage planet inflation could be responsible for the observed properties of this system.

\end{sciabstract}

\section*{One Sentence Summary:} An outlier among outliers, this discovery suggests hot Neptunes can retain atmospheres much longer than previously thought.

% be inflated by the evolution of their host star.

%Evolution models suggest this planet should have lost its atmosphere billions of years ago, but it has not. %

%The very low density and evolved state of this planet suggests hot Neptunes may retain their atmospheres via re-inflation.

%An outlier among outliers, this planet's existence suggests hot Neptunes can experience re-inflation.

%

\section*{Main Manuscript}

Over the last 30 years, the discovery of over 5,000 exoplanets has revealed that planetary system architectures are incredibly diverse \cite{berger2020,rosenthal2021}. One of the most striking demographic features of this sample is the dearth of planets with masses and radii similar to Neptune on short periods \cite{mazeh2016}. Star-planet interactions, including atmospheric effects like inflation \cite{guillot1996} and photoevaporation \cite{vidalmadjar2004} and orbital processes like high-eccentricity migration \cite{matsakos2016,owen2018}, have been proposed to be responsible for this feature of planet demographics.

The Transiting Exoplanet Survey Satellite (\emph{TESS}) has successfully confirmed more than 100 planets with orbital periods less than 10 days since 2018 \cite{guerrero2021}. Several of these planets have shown clear evidence for radius inflation \cite{rodriguez2021, yee2022, grunblatt2022}. Others are expected to have undergone significant atmospheric mass loss in their lifetimes \cite{jenkins2020,persson2022}. As inflated planets are more susceptible to mass loss, a process that continues over the lifetime of the planet, the characterization of inflated planets around post-main sequence, evolved (T$_{\mathrm{eff}} <$ 6000 K, R $>$ 2 R$_\odot$) stars nearing the end of their lives can constrain timescales for planetary mass loss and inflation mechanisms. Evidence for extended atmospheres of inflated planets transiting evolved stars has already been identified \cite{mounzer2022}.

TIC 365102760 is an evolved star at the base of the red giant branch that was observed by \emph{TESS} between 2019-Jul-18 and 2022-Sep-30. An initial box least squares (BLS) analysis \cite{kovacs2002} of the \texttt{giants} light curve revealed a planet candidate with orbital period 4.2 days and radius of $\approx$6.5 R$_\oplus$. We then flagged this planet candidate as a Community TESS Object of Interest (CTOI) on 2021-05-13. We determined a stellar mass of \starmass, a stellar radius of \starradius, and an age of 7.2 $\pm$ 1.4 Gyr for this system using a combination of photometric, astrometric, and spectroscopic data sets (see Supplementary Materials for details). A subsequent joint analysis of the radial velocity and \texttt{giants} light curve data using the \texttt{exoplanet} software package \cite{exoplanet} and our determinations of stellar parameters reveals a mass of \planetmass\ and radius of \planetradius\ for TIC 365102760 b, confirming the planetary nature of these signals (Figure 1, Table \ref{table:planet}). 

We find that based on its mass, radius and orbital period, \planet\ is a member of the relatively rare hot Neptune planet population \cite{jenkins2020,armstrong2020,persson2022}. Using the fitted values of \planetmass\ and \planetradius\ as the mass and radius for this planet, we determine a planet density of 0.437$^{+0.229}_{-0.150}$ g cm$^{-3}$, among the lowest densities yet measured for a hot Neptune. We also use our constraints on stellar mass, metallicity, and luminosity, and use MIST evolutionary tracks \cite{choi2016} to determine the current incident flux and incident flux history of this planet, assuming the planet has not migrated since the star reached an age of 20 Myr. We find that this planet is one of fewer than five currently known Neptunian (3-8 R$_\oplus$) planets orbiting post-main sequence, or evolved (T$_{\mathrm{eff}} <$ 6000 K, R $>$ 2 R$_\odot$) stars, and is the only hot Neptune (3-8 R$_\oplus$, $P$ $<$ 10 d) known in an evolved system (see Figure 2). 

We then use previously established relations between stellar bolometric flux and XUV flux as a function of stellar age \cite{king2021} to estimate the current XUV flux as well as the XUV flux history of \planet\ (for more details, see Supplementary Materials). We use these XUV flux estimates to determine the instantaneous rates of mass loss from this planet over its lifetime using a modified version of the ``energy-limited mass loss" equation \cite{watson1981, caldiroli2022}:

\begin{equation}
\dot{M} = \eta_\mathrm{eff} \frac{3F_{\mathrm{XUV}}}{4KG\rho_p} ,
\end{equation}
where $\dot{M}$ represents the planetary mass loss rate, $\eta_\mathrm{eff}$ is the efficiency of XUV evaporation from the planet, $F_{\mathrm{XUV}}$ is the stellar flux in high-energy, X-ray and ultraviolet wavelengths, $G$ is the gravitational constant, and $\rho_p$ is the mean planetary mass density. The correction factor $K$ which accounts for the fact that atmospheric compounds only need to reach the planet's Roche lobe radius to escape the planetary atmosphere \cite{erkaev2007}:

\begin{equation}
K = 1 - \frac{3}{2}\Bigg
(\frac{R_p}{R_\mathrm{Roche}}\Bigg) + \frac{1}{2}\Bigg(\frac{R_p}{R_\mathrm{Roche}}\Bigg)^3 ,
\end{equation}
where, following the small planet-to-star mass ratio approximation, 

\begin{equation}
R_\mathrm{Roche} \approx a\Bigg(\frac{M_p}{3M_*}\Bigg)^{1/3} . 
\end{equation}

We estimated $\eta_\mathrm{eff}$ as a function of $F_{\mathrm{XUV}}$ at each time step following the analytical approximation of \cite{caldiroli2022}, allowing us to determine instantaneous mass loss rates $\dot{M}$ for all of the planets in our sample. We then integrate these instantaneous mass loss rates over the lifetime of these planets using MIST stellar evolutionary tracks to determine the total fraction of mass lost by \planet\ over its lifetime relative to other known planetary systems, assuming planet formation and migration is complete after 20 Myr of stellar evolution. We compare the expected cumulative fraction of mass lost as a function of planet radius for systems with ages and well-characterized planet masses and radii in Figure 3. 

Using this formulation for atmospheric mass loss, we find that approximately 65\% of the current planet mass should have been lost over its lifetime (assuming that the planet radius has not changed over time), in good agreement with the fraction of mass loss inferred through an analytic approximation \cite{vissapragada2022b}. 

In order to confirm this cumulative atmospheric loss, we also determine the energy-limited mass loss rate using a more simplified equation taken from \cite{watson1981}:

\begin{equation}
\dot{M} = \frac{\varepsilon \pi R_{\mathrm{XUV}}^2 F_{\mathrm{XUV}}}{KGM_p/R_p} ,
\end{equation}

where $\varepsilon$ represents an XUV heating efficiency (we conservatively assume 0.1 \cite{shematovich2014, salz2016}), $R_\mathrm{XUV}$ is the effective absorption radius of the planet in XUV, and the other variables are the same as defined above. We assume an effective XUV absorption radius R$_\mathrm{XUV}$ = 1.1 R$_p$ based on previous estimates \cite{murrayclay2009}, which is likely an underestimate for this planet given its relatively low density and gravitational potential \cite{salz2016, krenn2021}. Using this conservative approach, we determine a cumulative mass loss fraction of 24\% for this planet.

Though the actual fraction of mass loss is difficult to estimate accurately \cite{paiasnodkar2022}, our predictions suggest that no currently known Neptune-sized or larger planets are expected to have experienced higher fractional amounts of mass loss than \planet. Overall, only three well-characterized planets with measured ages are expected to have experienced more fractional mass loss than \planet \cite{frustagli2020,weiss2021,serrano2022}. All of these planets are rocky, and thus any primordial atmosphere that might have existed has likely been stripped from them. Interestingly, although Jupiter-sized planets are predicted to be more resilient to mass loss than Neptune-sized planets \cite{owen2017, caldiroli2022}, no well-characterized Jupiter-sized exoplanets with measured ages listed on the NASA Exoplanet Archive are expected to have undergone such a high fraction of atmospheric mass loss as \planet. 

%The actual total mass loss of these rocky planets is expected to be much lower than predicted by the equations used here, as the atmospheric mass fraction is $<$1\% of the planetary mass, and the rocky component of the planet is left intact. 

% Conversely, for the few gaseous, Jupiter-sized planets expected to experience similarly high fractions of atmospheric mass loss, over 90\% of the planet's mass can be lost yet the planet radius can remain effectively unchanged, due to the wide range of hot Jupiter densities. The smaller gravitational potential of Neptune-mass planets likely makes them less resilient to atmospheric mass loss than hot Jupiters \cite{owen2017,caldiroli2022}. 

%hartman2016, west2016, demangeon2018, lam2021, pepper2017,

Models predict that $\approx$20-30\% of TIC365102760 b's mass should be contained in its gaseous envelope based on its incident flux, mass and radius \cite{lopez2014, chen2016}. Thus, assuming that the planet did not experience migration or inflation after a system age of 20 Myr, most or all of the planet's atmosphere should have been stripped over its lifetime \cite{chen2016}. However, as the planet's mass and radius imply it must have a gaseous envelope today, the planet may have had a much larger atmospheric mass in the past (and thus a much larger atmospheric mass fraction, implying higher fractional mass loss). Such a large atmospheric mass would also imply a planetary metallicity and core mass much lower than expected for a planet in this mass range \cite{thorngren2016, lee2019}. %Alternatively, this planet may have avoided significant atmospheric mass loss despite its current size and orbit.

Alternatively, \planet\ may have avoided such a high fraction of atmospheric mass loss through a number of scenarios, the simplest of which we consider here. First, the stellar flux in XUV may be significantly lower or absorbed less efficiently than existing models predict \cite{owen2019, king2021}, preventing severe atmospheric erosion even if the planet has not changed its orbit or radius since formation. We note that we find a stellar spectral activity measure log (R'$_\mathrm{HK}$) = -5.4 for our high-resolution spectroscopic observations of \hoststar, marginally lower than similar intermediate-mass, evolved planet host stars. XUV irradiation from an intermediate-mass star such as \hoststar\ is expected to be over an order of magnitude stronger than the Sun, larger than the difference in XUV flux on the main sequence for the models used here, but the range in measured activity level for intermediate-mass main sequence planet hosts spans more than an order of magnitude \cite{fossati2018, france2018}. Correlations between stellar activity levels and expected atmospheric erosion of other planets could help to validate this hypothesis. Mass loss could also be less efficient than models predict in this planet due to metal line cooling \cite{owen2012}, magnetic suppression \cite{owen2014}, or suppression via stellar winds \cite{wang2021}, although these effects are not expected to dominate the mass loss rate for the stellar mass and planet mass present in this system.

Second, the planet may have migrated to its current orbit during the main sequence lifetime of its host star from a previous larger orbit, avoiding the highest intensity of XUV irradiation from its host star. Both star-planet and planet-planet interactions could result in orbit reconfiguration. However, the absence of any transit timing variations, astrometric noise \cite{gaia2018} or additional radial velocity signal or trend suggests that there are no other planets relatively near to \planet\ in this system, suggesting that recent planet-planet interactions are not likely. Furthermore, \planet\ does not appear to have a high-eccentricity orbit, suggesting that migration due to star-planet interaction is also unlikely or not very recent in the system's history \cite{villaver2014}. Better constraints on the eccentricity via additional radial velocity followup, or constraint of the obliquity of the planetary orbit relative to the stellar spin axis could also support or refute evidence for orbital migration \cite{bourrier2023}.

Finally, the planet may have been significantly smaller in the past, limiting the instantaneous rate of mass loss on the main sequence. Our estimates of cumulative mass loss assume no change in planet radius as a function of time, which is not expected for planets in this mass and temperature regime experiencing significant mass loss \cite{thorngren2023}. Atmospheric mass loss tends to shrink planets  over time in this mass and temperature regime, resulting in more mass loss than what we assume in Figure 3. However, if we assume the recent increase in irradiation received by this planet due to post-main sequence evolution could have resulted in rapid re-inflation of this planet \cite{thorngren2021}, this implies a smaller radius for the planet when \hoststar\ was on the main sequence.

%We develop two toy models of the planet radius evolution over time of \planet\ assuming rapid re-inflation and atmospheric mass loss.
 
Following the formulation of \cite{thorngren2021} and assuming a late-stage inflation efficiency comparable to what has been observed in other systems \cite{grunblatt2017}, we can model changes in the radius of \planet\ over the lifetime of the system. We compare this to an atmospheric mass loss model of the planet radius following \cite{thorngren2023} assuming an initial planet mass of 29.2 M$_\oplus$ and planet metallicity fraction of 0.8. We show the expected radius evolution as a function of time for both of these scenarios using the observed planet mass, radius and age in Figure 4. We note that at 1 Gyr, the inflated planet model radius is $\approx$16\% smaller than the current radius, while the atmospheric mass loss model radius is $\approx$5\% larger. As the atmospheric mass loss rate depends on the density of the planet, this implies a $\sim$50\% reduction in instantaneous mass loss at that time for the late-stage inflation model relative to the atmospheric mass loss model. Since stellar XUV irradiation is maximized at ages of 1 Gyr or less, this implies a similar $\sim$50\% reduction of total atmospheric mass loss of the planet if it has inflated during post-main sequence evolution. 

% Post-main sequence inflation has been observed in other evolved planetary systems \cite{grunblatt2017, wittenmyer2022}.

Current observational evidence for both late-stage inflation and/or weak photoevaporation is stronger than evidence for late-stage migration in this system. As our toy model of planet radius inflation would still result in more atmospheric mass loss than the total atmospheric mass expected for a planet of this temperature and radius, and the XUV flux of intermediate-mass stars is predicted to be stronger than what was inferred here, this implies both a lower-than-expected level of XUV irradiation and late-stage inflation contribute to the continued existence of this planet's atmosphere. It is also possible that the planet both lost a significant amount of mass during the main sequence via photoevaporation and subsequently became inflated during post-main sequence evolution.

% Assuming a re-inflation efficiency comparable to what has been predicted in other systems \cite{lopez2016, grunblatt2016,grunblatt2017}, we can predict the main sequence radius of TIC 365102760 b assuming a similar rate of re-inflation. If this planet increased in radius by a similar rate as predicted for other re-inflated planets  ($\approx$30\%), this would reduce overall mass loss of the planet by a factor of $\sim$2, allowing the planet to hold onto a much larger fraction of its atmosphere until the present day. As the mass of this planet is significantly smaller and the temperature significantly higher for \planet\ than previous re-inflation candidates, we expect that re-inflation processes would produce a larger change in radius for this planet than previously reported for re-inflation candidates. Thus, we posit that the observed state of this planet may be evidence for planetary re-inflation around post-main sequence stars at both Jupiter-like and Neptune-like planet masses. It is also possible that the planet both lost a significant amount of mass during the main sequence via photoevaporation and subsequently became re-inflated during post-main sequence evolution.

The discovery of a low-density hot Neptune orbiting an evolved star demonstrates that the atmospheres of these planets are more resilient than previously thought. Furthermore, it demonstrates that planets that are smaller than Jupiter in size may be inflated directly by irradiation from their host stars. This has important implications for understanding the structure and evolution of Neptune-sized planets, and interpreting the demographics of the known planet population. Finding more evolved, hot, Neptune-sized planets at different masses and densities may reveal additional trends with planet composition or stellar activity. Focused searches for these evolved systems are necessary as these planets are missed by general searches for transiting planets (e.g., \cite{grunblatt2023}).  Additional observations of \planet\ using ground-based and space-based spectroscopic approaches may reveal atmospheric outflows from this planet, better constraining the lifetime of its atmosphere as well as its atmospheric composition. An extended atmosphere potentially indicative of outflow has already been detected via sodium D line transmission in another evolved, inflated planet, KELT-11 b \cite{mounzer2022}, though at levels lower than initially expected. This may be due to high-altitude clouds, which may be related to silicate cloud production resulting from rapid planet re-inflation \cite{gao2020,thorngren2021}. Constraining the balance between planet atmospheric inflation and mass loss will help reveal the evolution of planetary atmospheres over time, clarifying planet demographic features such as the hot Neptune desert.

\bibliography{scibib}

\bibliographystyle{Science}

% Following is a new environment, {scilastnote}, that's defined in the
% preamble and that allows authors to add a reference at the end of the
% list that's not signaled in the text; such references are used in
% *Science* for acknowledgments of funding, help, etc.

% \begin{scilastnote}
% \item We've included in the template file \texttt{scifile.tex} a new
% environment, \texttt{\{scilastnote\}}, that generates a numbered final
% citation without a corresponding signal in the text.  This environment
% can be used to generate a final numbered reference containing
% acknowledgments, sources of funding, and the like, per {\it Science\/}
% style.  Along those lines, we'd like to thank readers of this document
% for their attention, and invite them to address any questions to
% Stewart Wills, at swills@aaas.org.
% \end{scilastnote}

% For your review copy (i.e., the file you initially send in for
% evaluation), you can use the {figure} environment and the
% \includegraphics command to stream your figures into the text, placing
% all figures at the end.  For the final, revised manuscript for
% acceptance and production, however, PostScript or other graphics
% should not be streamed into your compliled file.  Instead, set
% captions as simple paragraphs (with a \noindent tag), setting them
% off from the rest of the text with a \clearpage as shown  below, and
% submit figures as separate files according to the Art Department's
% instructions.

\clearpage

\section*{Acknowledgements}
We acknowledge Kaz Gary, Zafar Rustamkulov, Eric Lopez, Nestor Espinoza and Ji Wang for helpful conversations that helped to improve this manuscript. We acknowledge the use of public TESS data from pipelines at the TESS Science Office and at the TESS Science Processing Operations Center. Resources supporting this work were provided by the NASA High-End Computing (HEC) Program through the NASA Advanced Supercomputing (NAS) Division at Ames Research Center for the production of the SPOC data products. This work was supported by a NASA Keck PI Data Award, administered by the NASA Exoplanet Science Institute. Data presented herein were obtained at the W. M. Keck Observatory from telescope time allocated to the National Aeronautics and Space Administration through the agency's scientific partnership with the California Institute of Technology and the University of California. The Observatory was made possible by the generous financial support of the W. M. Keck Foundation. The authors wish to recognize and acknowledge the very significant cultural role and reverence that the summit of Maunakea has always had within the indigenous Hawaiian community. We are most fortunate to have the opportunity to conduct observations from this mountain. This research has made use of the Exoplanet Follow-up Observation Program website, which is operated by the California Institute of Technology, under contract with the National Aeronautics and Space Administration under the Exoplanet Exploration Program. 

\section*{Funding}
S.G., N.S., and D.H. acknowledge support by the National Aeronautics and Space Administration under Grant 80NSSC19K0593 issued through the TESS Guest Investigator Program. D.H. acknowledges support from the Alfred P. Sloan Foundation, the National Aeronautics and Space Administration (80NSSC21K0652), and the Australian Research Council (FT200100871). N.S. acknowledges support from the National Science Foundation through the Graduate Research Fellowship Program under Grants 1842402 and DGE-1752134. S.V. acknowledges support from the Heising-Simons Foundation under Grant 2022-3578. M.R. acknowledges support from the Heising-Simons Foundation under Grant \#2022-3538. J.M.A.M. is supported by the National Science Foundation Graduate Research Fellowship Program under Grant No. DGE-1842400. Any opinions, findings, and conclusions or recommendations expressed in this material are those of the authors and do not necessarily reflect the views of the National Science Foundation. Funding for the TESS mission is provided by NASA's Science Mission Directorate. 

\section*{Author contributions}
S.K.G. wrote all text and generated all figures in the manuscript. S.K.G. and D.H. jointly designed the overall target selection for this planet search, and S.K.G. and N.S. jointly designed the \texttt{giants} pipeline in order to identify candidates. S.K.G. identified \hoststar\ as a potential planet candidate. S.G., M.M., A.C., E.T., C.B., J.M.A.M., M.R., H.I., and A.H. acquired the radial velocity observations. D.H. and R.A. helped oversee the initial planet search. K.S. performed the stellar analysis used to characterize the fundamental properties of \hoststar\ and KELT-11. S.Y., S.V., and D.T. assisted with the interpretation of the planet discovery and its implications for the history of this planetary system.

\section*{Competing interests}
We declare no competing interests.

\section*{Data and materials availability}
The \texttt{giants} pipeline is available at www.github.com/nksaunders/giants, from which the \texttt{giants} light curve can be reproduced. TESS Full Frame Image data and QLP and SPOC lightcurves can be downloaded from the publicly available MAST Archive. Keck/HIRES measurements will be made public via the Keck Observatory Archive 18 months after observations were taken.

\clearpage

\begin{figure*}[h!]
    \centering
    \includegraphics[width=0.65\textwidth]{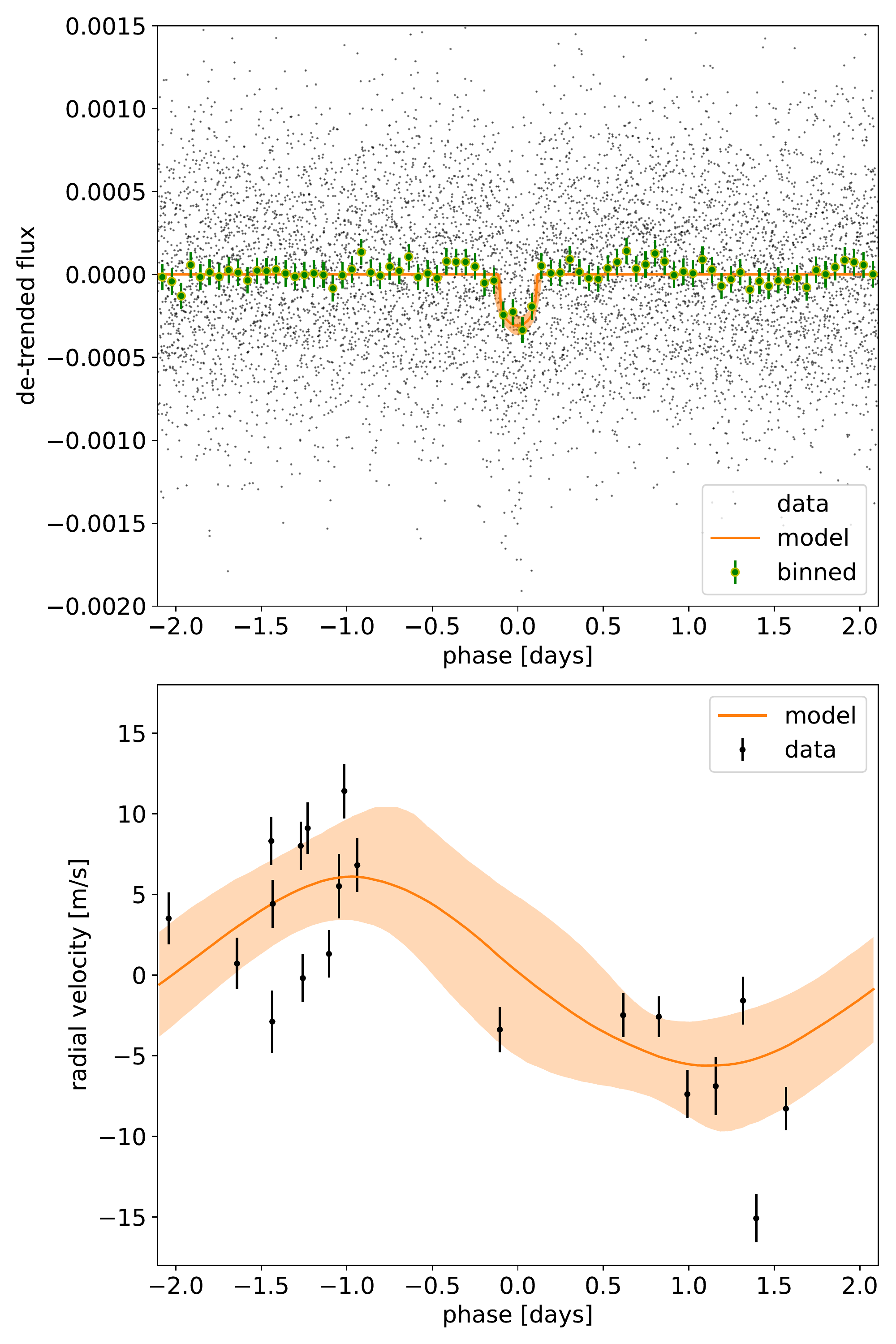}
    \label{fig:lc_rv_3651}
\end{figure*}
\vspace{-0.3cm}
\noindent {\bf Fig. 1.} {\it Top:} The \texttt{giants} light curve of \hoststar\ folded at a period of \period. The de-trended photometry is shown in black with the binned photometry plotted in green and best fit \texttt{exoplanet} model overplotted in orange, with 90\% confidence intervals shaded. {\it Bottom:} All radial velocity observations of \hoststar\ (black) along with the best fit \texttt{exoplanet} model and 90\% confidence interval (orange) folded at the orbital period of the planet. Observations come from the Keck-I/HIRES spectrograph on Maunakea.

\clearpage

\begin{figure*}[ht!]
    \centering
    \includegraphics[width=\textwidth]{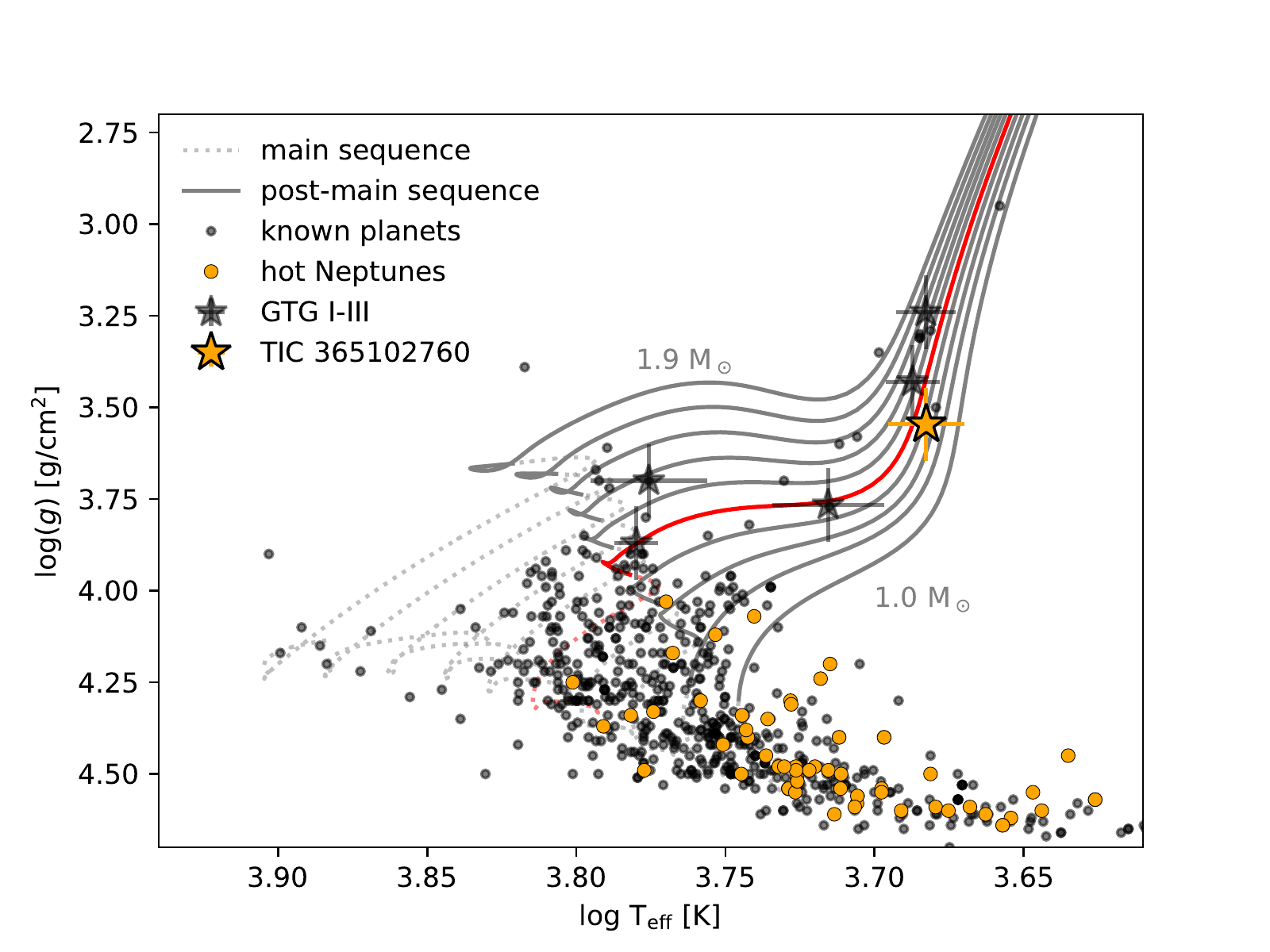}
    \label{fig:hrdiag}
\end{figure*}

\noindent {\bf Fig. 2.} Surface gravity versus effective temperature displayed logarithmically for \hoststar\ along with the other planets discovered by the \emph{TESS} Giants Transiting Giants (GTG) program (stars) compared to all known planets with well-characterized radii and masses (points). GTG discoveries have almost doubled the number of known systems on the subgiant and red giant branch. Hot Neptunes (3 R$_\oplus$ $<$ R$_p$ $<$ 8 R$_\oplus$, P$_\mathrm{orb}$ $<$ 10 d) have been highlighted in orange. We also illustrate MIST evolutionary tracks of 1-2 M$_\odot$, +0.25 [Fe/H] dex stars in 0.1 M$_\odot$ increments for reference. We have highlighted a MIST evolutionary track for a 1.4 M$_\odot$, [Fe/H] = 0.25 dex star in red, illustrating the rough evolutionary sequence probed by the GTG survey.

% Please do not use figure environments to set
% up your figures in the final (post-peer-review) draft, do not include graphics in your
% source code, and do not cite figures in the text using \LaTeX\
% \verb+\ref+ commands.  Instead, simply refer to the figure numbers in
% the text per {\it Science\/} style, and include the list of captions at
% the end of the document, coded as ordinary paragraphs as shown in the
% \texttt{scifile.tex} template file.  Your actual figure files should
% be submitted separately.

\clearpage

\begin{figure*}[ht!]
    \centering
    \includegraphics[width=.99\textwidth]{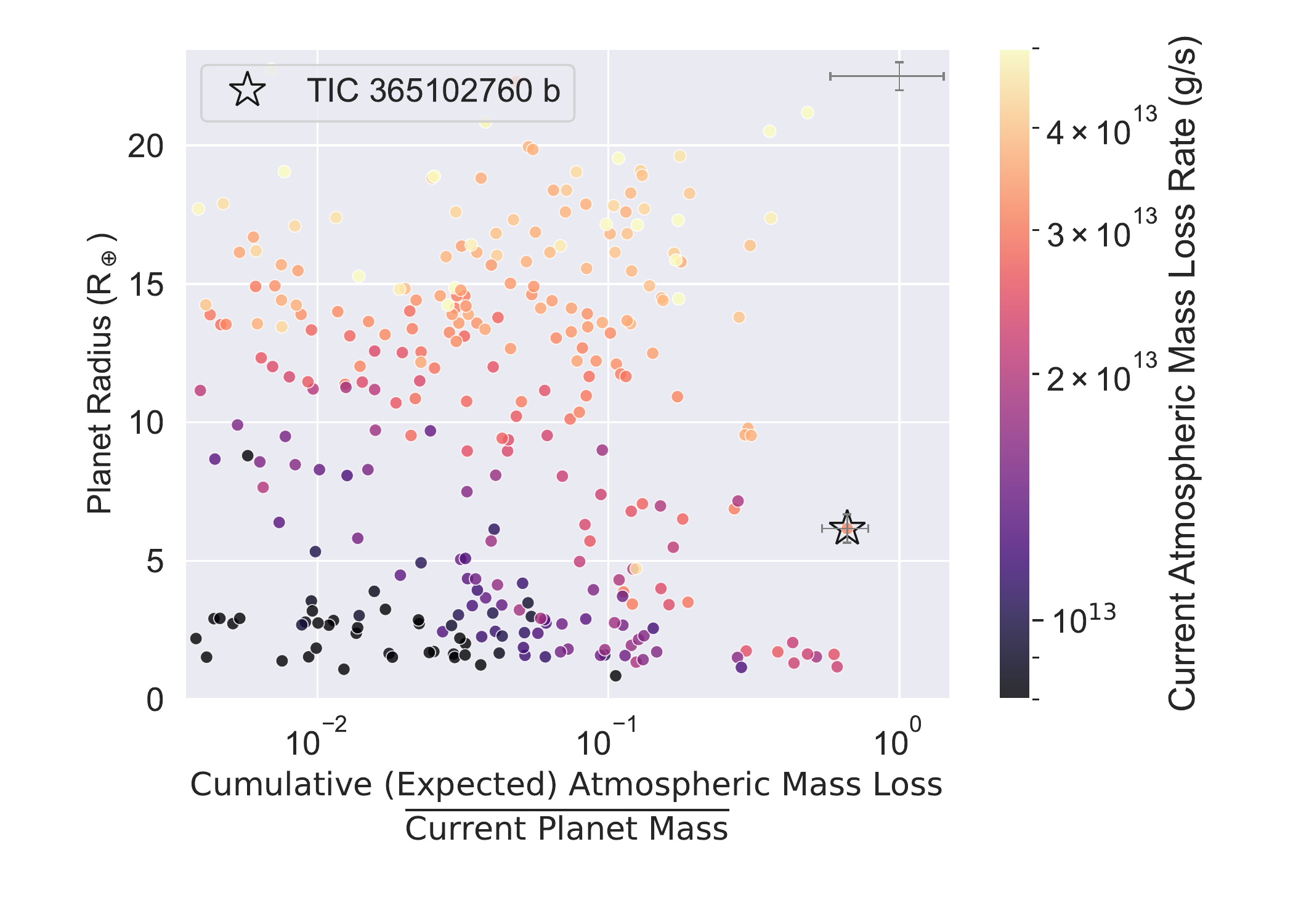}
    \label{fig:fracmassloss}
\end{figure*}

\noindent {\bf Fig. 3.} Planet radius versus the fraction of estimated total atmospheric mass loss divided by the current planet mass for all planets with masses and radii reported with $\geq$50\% accuracy, and ages on the NASA Exoplanet Archive accessed on 31-Jan-2023. Color indicates atmospheric mass loss rate in grams/second. Evolved planets have been highlighted with stars, where \planet\ is indicated by the largest star. Average errors are illustrated in the upper right hand corner of the figure.  \planet\ both has a higher current mass loss rate and is predicted to have experienced more atmospheric mass loss than all other gaseous planets, with only a few rocky planets experiencing higher rates of mass loss. As the atmospheric mass for this planet is predicted to be smaller than the total mass expected to have been lost, it is unclear how \planet\ has retained an atmosphere given its current radius and orbit around its host star.%, unless the host star has been atypically low in XUV irradiation throughout its lifetime, and/or the planet has undergone radius or orbit evolution since its formation.

\clearpage

\begin{figure*}[ht!]
    \centering
    \includegraphics[width=.99\textwidth]{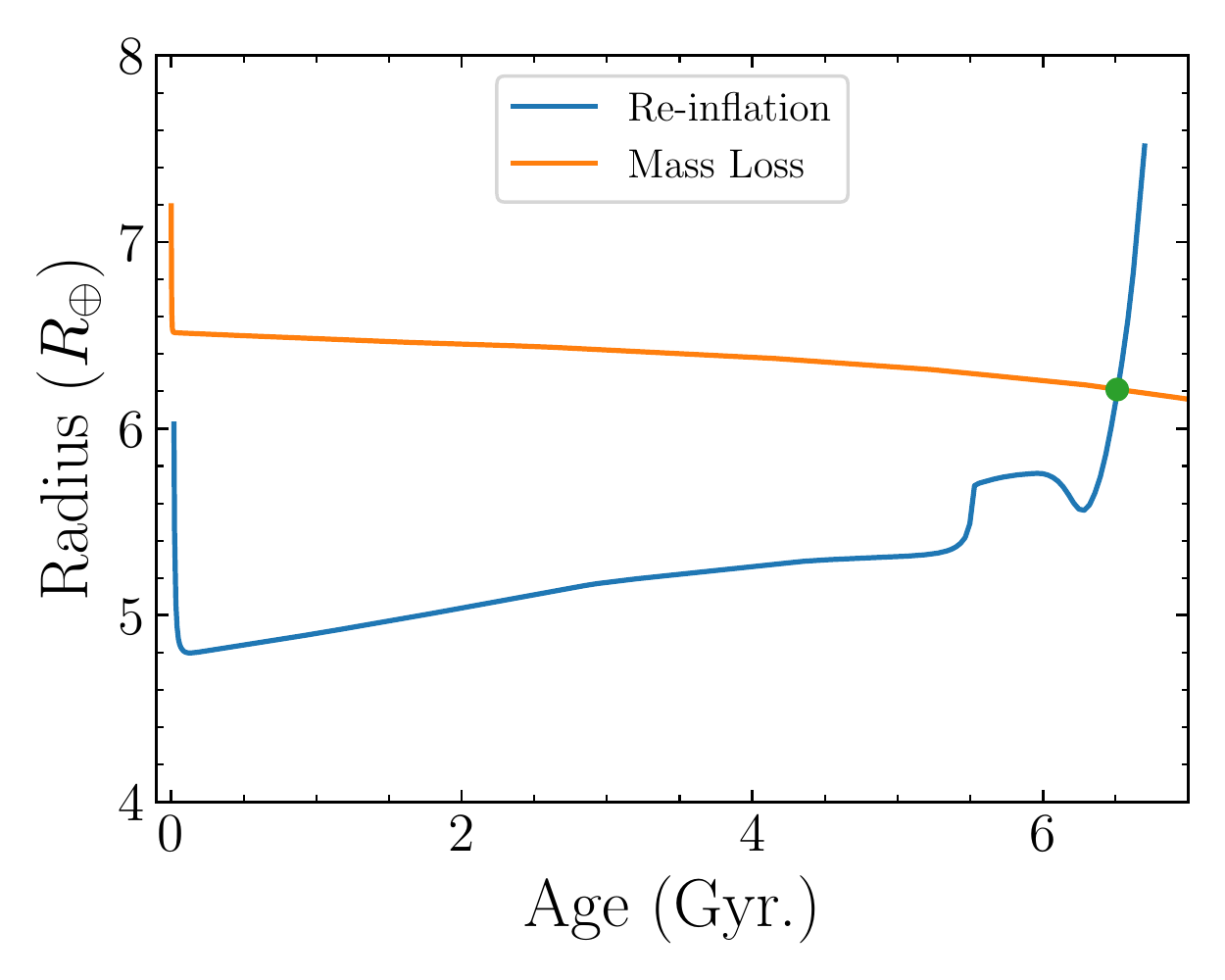}
    \label{fig:radevol}
\end{figure*}

\noindent {\bf Fig. 4.} Planet radius versus age for two toy models of \planet\ that match the median mass, radius, age and flux measured here (green point). A planet model following the rapid re-inflation formulation of \cite{thorngren2021} is shown in blue, while an atmospheric mass loss model following the formulation of \cite{thorngren2023} assuming an initial planet mass of 29.2 M$_\oplus$ and planet metal composition fraction of 0.79 is shown in orange. This rapid re-inflation model implies the planet was smaller during the main sequence phase, resulting in a $\sim$50\% reduction in total atmospheric mass loss from \planet\ relative to models that do not allow late-stage inflation.%, unless the host star has been atypically low in XUV irradiation throughout its lifetime, and/or the planet has undergone radius or orbit evolution since its formation.

% {\it Left:} Orbital period versus planet radius for confirmed planets and new candidates transiting evolved (R$_*$ $>$ 2 R$_\odot$, T$_\mathrm{eff}$ $<$ 6000 K) stars. Planets known around evolved stars before the launch of \tess are shown in green. Those confirmed by \tess\ are shown as the largest symbols. Additional community-flagged planet candidates found by \tess\ are shown as small red stars, and \tess\ Objects  of Interest (TOIs) are shown in gray. \planet\ is shown by the largest star near the center of the Figure, and is the smallest planet found so far around an evolved star. {\it Right:} Orbital period versus planet radius for confirmed planets and new candidates transiting evolved (R$_*$ $>$ 2 R$_\odot$, T$_\mathrm{eff}$ $<$ 6000 K) stars. Planets known around evolved stars before the launch of \tess are shown in green. Those confirmed by \tess\ are shown as the largest symbols. Additional community-flagged planet candidates found by \tess\ are shown as small red stars, and \tess\ Objects  of Interest (TOIs) are shown in gray. \planet is shown by the largest star near the center of the Figure, and is the smallest planet found so far around an evolved star.

\clearpage

\section*{List of Supplementary Materials}

\setcounter{figure}{0}
\renewcommand{\figurename}{Fig.}
\renewcommand{\thefigure}{S\arabic{figure}}
\renewcommand{\thetable}{S\arabic{table}}

% \tableofcontents

Contents:

Materials and Methods

Supplementary Text

Figures S1-S10

Tables S1-S3

\clearpage

\section*{Materials and Methods}

\subsection*{Observations} \label{sec:methods}

\subsubsection*{\tess\ Photometry} \label{sec:photo}

\planet\ was discovered as part of the \tess\ Giants Transiting Giants survey identifying new planets around evolved host stars. Using version 8 of the \tess\ Input Catalog \cite{stassun2019}, we followed the cuts of \cite{grunblatt2019} based on color, magnitude, and Gaia parallax in order to limit our sample to evolved stars. We developed the \giants\footnote{https://github.com/nksaunders/giants} Python package for accessing, de-trending, and searching \tess\ observations for periodic transit signals. The details of how this pipeline processes \tess\ full frame image data are described in \cite{saunders2022}. 

\begin{figure*}
    \centering
    \includegraphics[width=\textwidth]{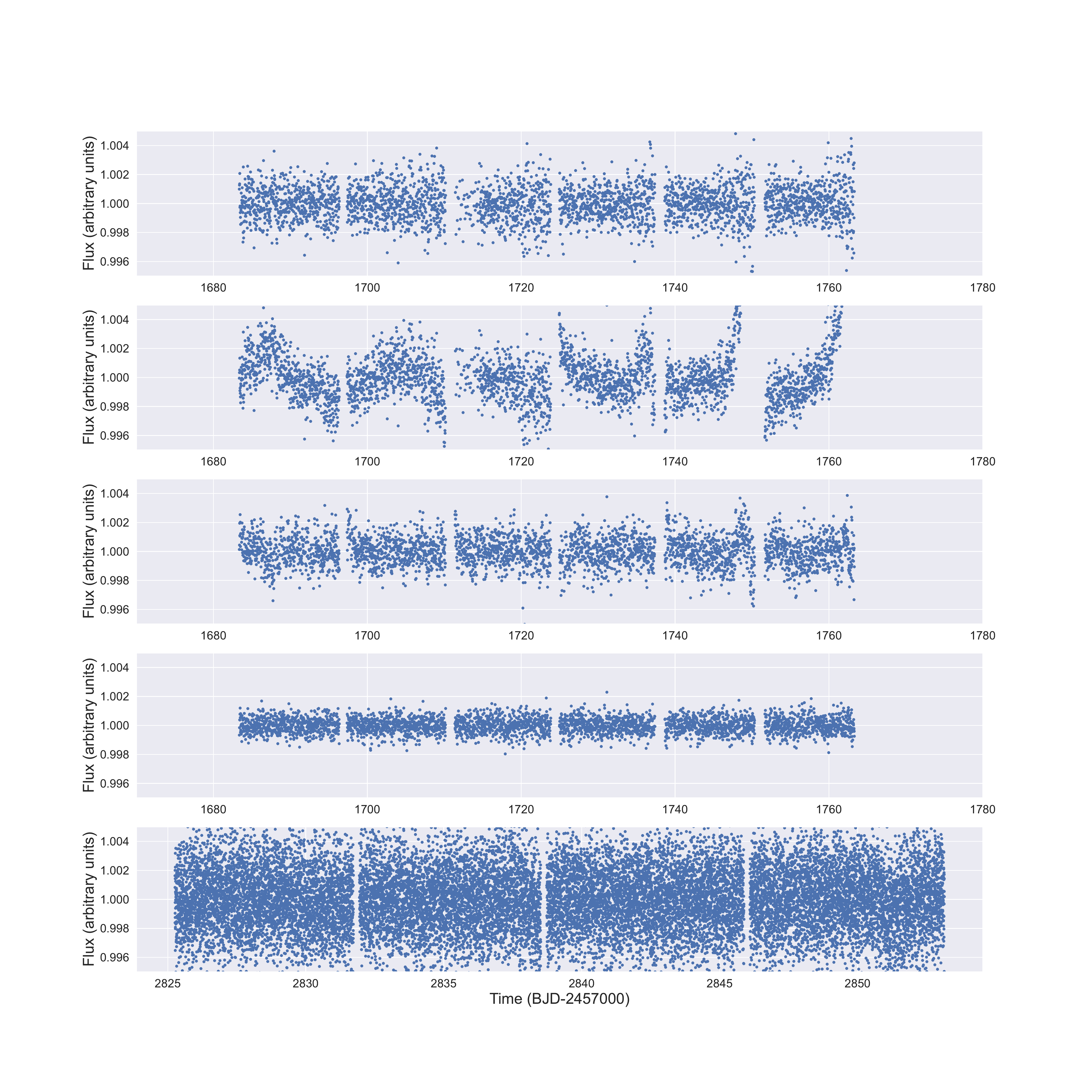}
    \caption{Full \tess\ prime mission light curves of \hoststar, produced using the QLP KSPSAP detrending, the QLP SAP detrending, the \texttt{eleanor} pipeline and the \texttt{giants} pipeline from top to second from bottom. Additional data have since been acquired by the \tess\ Extended Mission at higher cadence for \hoststar\ and converted into a light curve via the SPOC pipeline, plotted in the bottom row. The scatter is noticeably higher for the high-cadence light curve.}
    \label{fig:unfolded_lc}
\end{figure*}

\hoststar\ was observed in \emph{TESS} Sectors 14, 15, 16, 41, 55, and 56 between 2019-Jul-18 and 2019-Oct-07, 2021-Jul-23 to 2021-Aug-20, and 2022-Aug-05 to 2022-Sep-30. Optical photometry was first acquired at 30-minute cadence in 2019, 10-minute cadence in 2021, and 200-second and 2-minute cadence in 2022. We produced a light curve with both 30-minute and 10-minute cadence using the \texttt{giants} package. We also produced a light curve for this target using the \texttt{eleanor} package \cite{feinstein2019}. In addition, light curves for this target were produced by the \tess\ Science Team using both simple aperture photometry (SAP) as well as a Kepler-like signal processing algorithm (KSPSAP). Subsequently, the SPOC pipeline \cite{jenkins2010,jenkins2016}, which has been tested on a significantly larger number of planetary transit datasets than the other pipelines mentioned here, produced a light curve for the 120-second, high-cadence data, which we also use in our analysis. We present these five light curves for \hoststar\ in Figure \ref{fig:unfolded_lc}.

% \hoststar\ was observed in \emph{TESS} Sectors 14, 15, 16, 41, 55, and 56 between 2019-Jul-18 and 2019-Oct-07, 2021-Jul-23 to 2021-Aug-20, and 2022-Aug-05 to 2022-Sep-30. Optical photometry was first acquired at 30-minute cadence in 2019, 10-minute cadence in 2021, and 200-second and 2-minute cadence in 2022. The light curve of this target was extracted from the 30-minute, 10-minute, and 200-second cadence photometry using both the \texttt{giants} and QLP pipelines \cite{huang2020,saunders2022}, and using the SPOC pipeline for the 2-minute cadence photometry \cite{jenkins2016}. 

\begin{figure*}
    \centering
    \includegraphics[width=\textwidth]{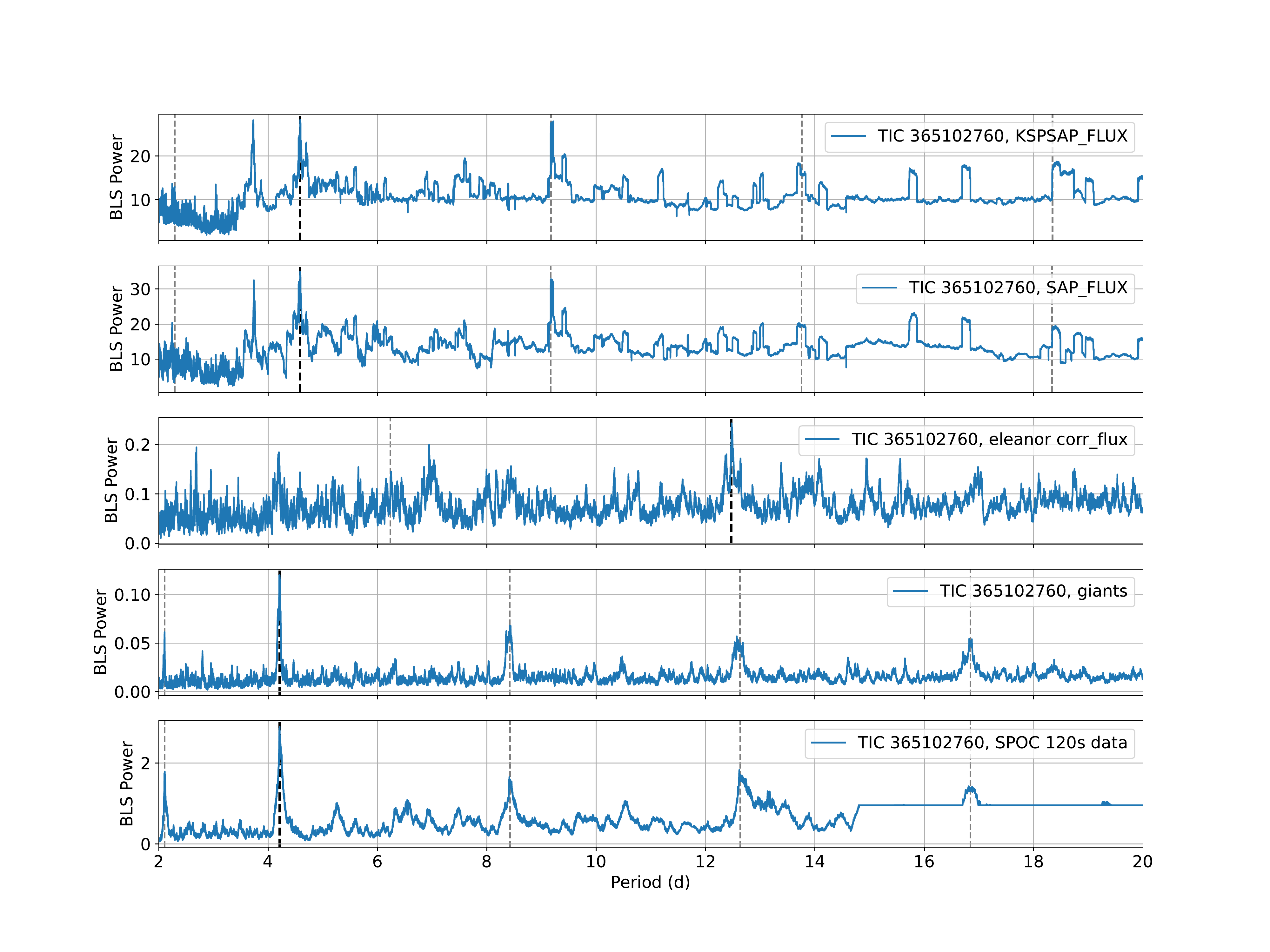}
    \caption{Box least squares power as a function of period for the five light curves of \hoststar\ shown in Figure S1. The period of highest power has been highlighted with a black dashed line, and its harmonics with gray dashed lines. The 4.2-day transit signal is only visible in the \texttt{giants} and SPOC light curves, while going undetected in the others.}
    \label{fig:bls_comp}
\end{figure*}

\begin{figure*}
    \centering
    \includegraphics[width=\textwidth]{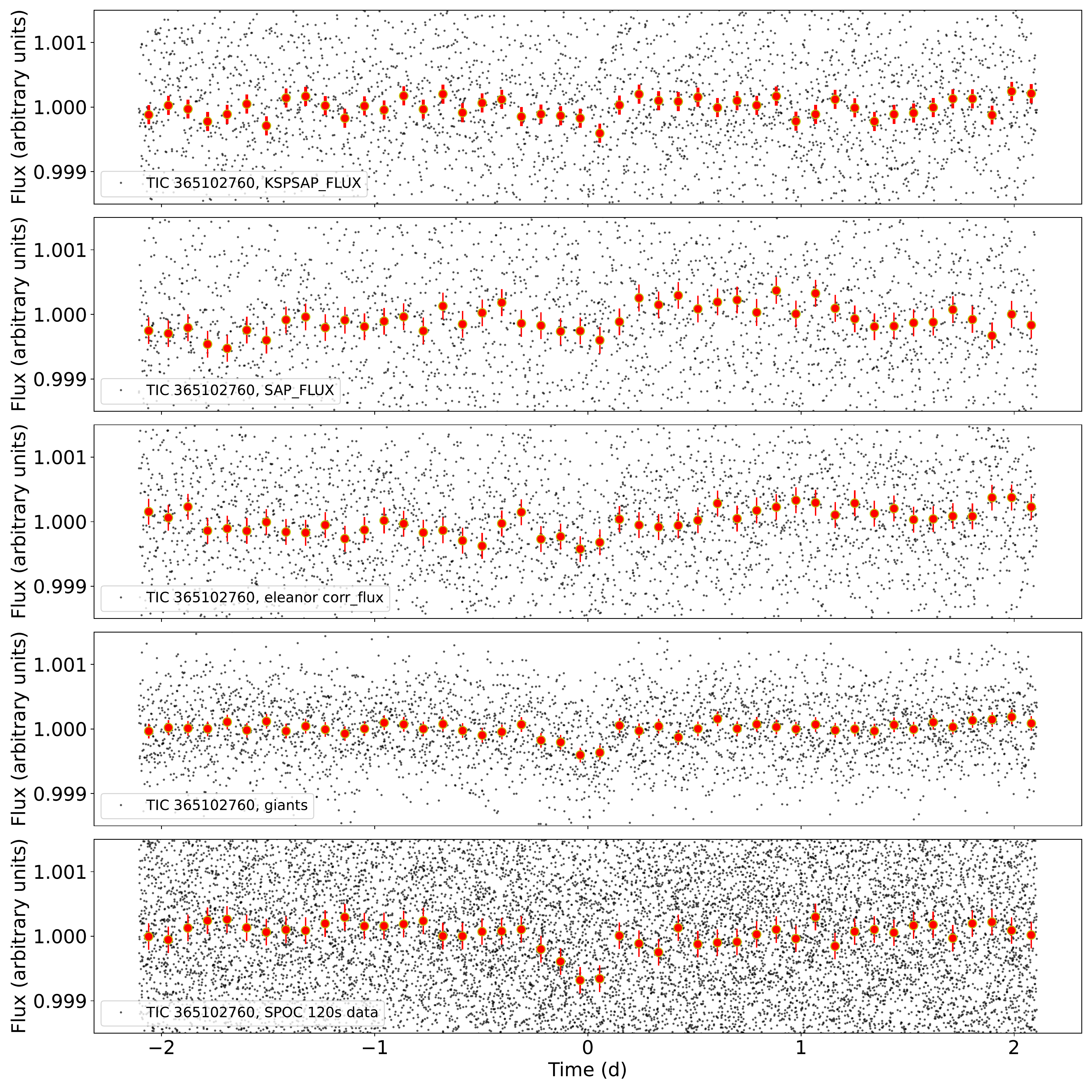}
    \caption{Phase-folded light curves of \hoststar\ using the QLP `KSPSAP FLUX', `SAP FLUX', \texttt{eleanor}, \texttt{giants}, and SPOC pipelines (from top to bottom, respectively). Clear differences in transit depth and shape and light curve scatter can be seen between the different light curves. The SPOC data covers a shorter baseline (only 27 days) but has 15x higher cadence than the other light curves.}
    \label{fig:folded_trans_comp}
\end{figure*}

In addition to \hoststar, we used our \texttt{giants} pipeline to produce \tess\ light curves for as many red giant branch stars with \tess\ magnitude $m_T <$ 13 as possible. We produced light curves for approximately 540,000 stars from the first 2 years of data from the \tess\ Mission. We performed an automated box least squared (BLS) \cite{kovacs2002} search on all targets, and produced summary plots displaying the BLS output as well as TIC information and the \tess\ full frame image pixel cut out. These summary plots were then visually inspected, during which this candidate was flagged for potential rapid ground-based follow-up. We illustrate the BLS results for all light curves of \hoststar\ considered in this work in Figure \ref{fig:bls_comp} and corresponding phase-folded light curves of \hoststar\ in Figure \ref{fig:folded_trans_comp}.

The 4.2-day transit signal from the planet is detected clearly above the noise floor in the BLS searches of the \texttt{giants} and 2-minute cadence SPOC light curves ($\geq$ $4\sigma$), but is not recovered in the other light curves. This is likely related to the principal component detrending as well as the median smoothing filter of the \texttt{giants} light curves, which is broader than that of the KSPSAP light curve but still results in a much smoother light curve than pure simple aperture photometry (as can be seen in Figure \ref{fig:unfolded_lc}). We note that in Figure \ref{fig:folded_trans_comp}, the transit can just barely be perceived by eye at a phase of 0.0 days in the \texttt{giants} and SPOC light curves, but is not visually detectable in the other light curves presented here. 

Given the relatively weak planet transit signal, this planet did not reach the \tess\ team criteria to be flagged as a \tess\ Object of Interest, and thus was not prioritized for ground-based follow-up by the larger \tess\ Follow-up Program. Thus, it was important for our team to independently verify the existence of a planetary signal around this star. %In order to do so, follow-up radial velocity measurements of this target were taken from the ground.

\subsubsection*{Radial Velocity Measurements}

Our team acquired RV observations of \hoststar\ with the HIRES spectrograph on the Keck-I telescope on Maunakea, Hawaii \cite{vogt1994}. HIRES has a resolving power of $R\approx120,000$ and wavelength coverage between $\sim350$nm and $\sim620$nm. 20 RV measurements of \hoststar\ were taken between June 13, 2021 and July 11, 2022. We list our RV measurements of \hoststar\ in Table \ref{table:rvs}.

\begin{figure*}
    \centering
    \includegraphics[width=0.9\textwidth]{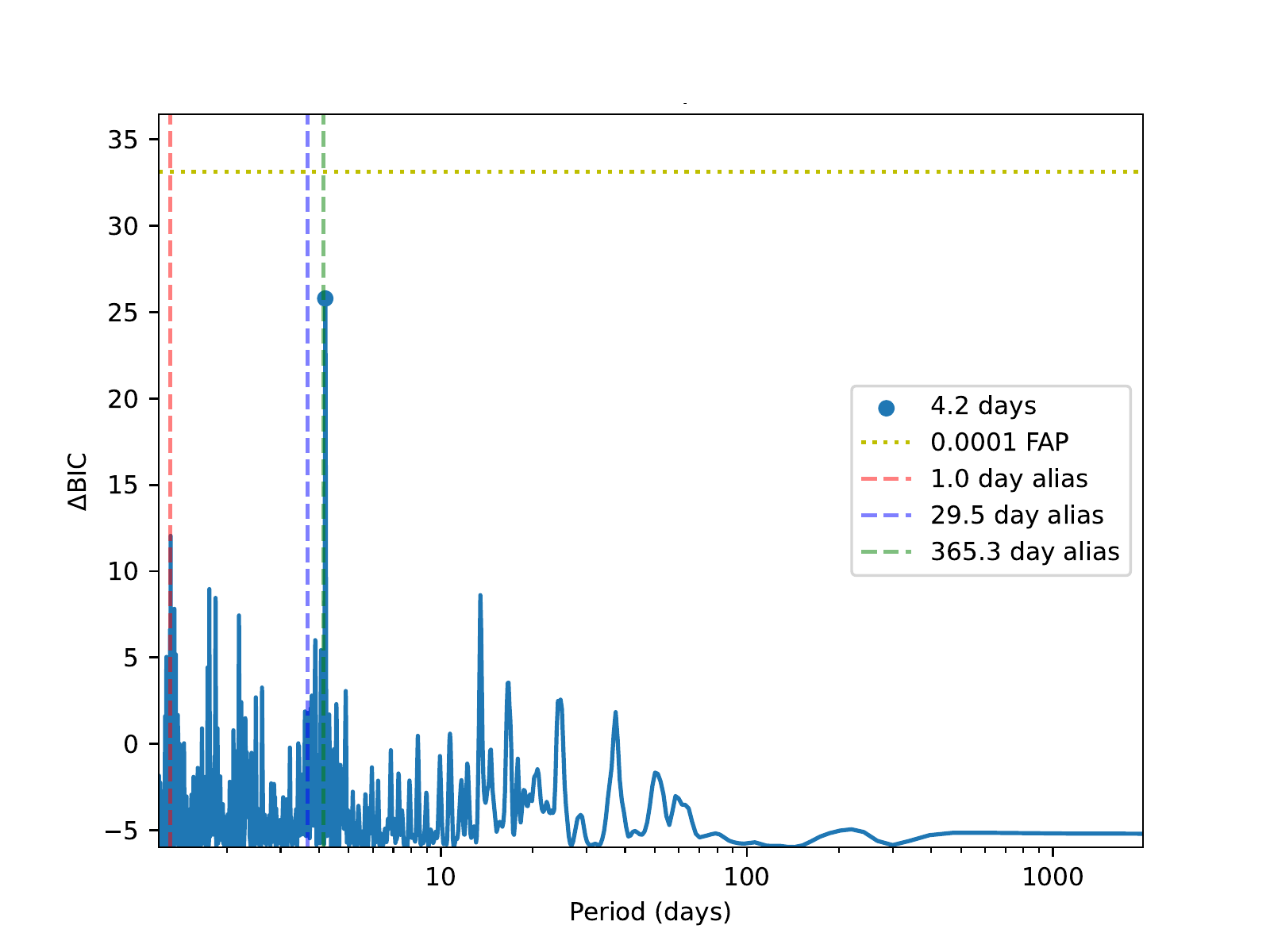}
    \caption{Periodogram of the Keck/HIRES radial velocity measurements of \hoststar. The strongest signal present in the data which does not correspond to a known alias of the Earth or Moon is highlighted with a blue point. We find a power excess in the radial velocity measurements at 4.2 days, with a $\Delta$BIC=25.5, just below the false alarm probability threshold of $\Delta$BIC=33 found to validate a planetary signal without additional information. This 4.2-day signal matches the period and phase of the signal found in the light curve of \hoststar.}
    \label{fig:rvsearch}
\end{figure*}

Using these RV measurements alone, we also search for an independent signal in the data. Using the software package \texttt{RVsearch} \cite{rosenthal2021} we convert the radial velocity measurements to a periodogram, which accounts for the systematic signals of the Earth and Moon orbital and rotational periods. We find that a signal at 4.20 days is recovered with high significance, matching the planet transit seen in the \texttt{giants} light curve. This signal falls just short of the false alarm probability cutoff used for RV-only planet confirmation. We illustrate detection of this signal in our RV data in Figure \ref{fig:rvsearch}. This independent confirmation of a planetary signal, observed with the same period and phase as seen in the \texttt{giants} light curve, makes the planet scenario significantly more likely than either dataset would on its own, and thus allows us to confirm this planet as real.

\subsubsection*{Visual Companions to \hoststar} 

While \hoststar\ does not have high resolution imaging results, the \emph{Gaia} astrometric noise metric RUWE is low (0.822), and the system is not included in the \emph{Gaia} eDR3 catalog of binary systems \cite{elbadry2021}. This indicates that the star is not in a binary system which could be resolved by \emph{Gaia} photometry, and implies no significant dilution in the \tess\ FFI-generated light curves. This is supported by the consistent transit depth seen in both the SPOC and \texttt{giants} light curves, where a dilution correction factor has been applied to the SPOC light curves only. Additionally, no evidence of a spectroscopic binary can be seen in the spectra of this star, placing limits on close stellar binarity. The radial velocity measurements of this system also do not show any significant linear or quadratic trends with time, suggesting \hoststar\ is a single star and not part of a binary system. 

% It is worth noting that although unlikely, if there are any close companion or background stars which are diluting the \tess\ light curve for this system, this would imply the planet radius is underestimated, and thus the planet is even lower density and larger radius than is reported here. This would cause atmospheric mass loss of this planet at a higher rate, making it more difficult for the planet to maintain its atmosphere until the current time. This could be interpreted as additional evidence for late-stage planet inflation.

\subsection*{Host Star Characterization} \label{sec:hoststar}

% \subsubsection*{Spectroscopic Analysis}

% We used SpecMatch to measure the metallicity, surface gravity and effective temperature of the host star from our HIRES template spectrum \cite{petigura2015}. We then used \textsf{isoclassify} \cite{huber2017} to combine TICv8 and spectroscopic information to determine stellar properties for the Giants Transiting Giants sample. Figure 2 shows an H-R diagram with evolutionary tracks downloaded from the MESA Isochrones \& Stellar Tracks (MIST; \cite{paxton2011,dotter2016,choi2016}). All Giants Transiting Giants systems were fit using \texttt{isoclassify} to highlight relative similarities between them. As all host stars have roughly the same mass and metallicity (M$_*$ $\approx$ 1.4 M$_\odot$, [Fe/H] $\approx$ 0.25 dex), we highlight this evolutionary track in red, and suggest that these systems may extend earlier analyses of ``retired A stars" to reveal evolutionary sequences for planetary systems orbiting post-main sequence, intermediate-mass stars \cite{johnson2007a,johnson2010}. We find that \hoststar\ is near the base of the red giant branch stage of evolution. 

\subsubsection*{Spectroscopic and Photometric Analyses}

We used SpecMatch to measure the metallicity, surface gravity and effective temperature of the host star from our HIRES template spectrum \cite{petigura2015}. We then used \textsf{isoclassify} \cite{huber2017} to combine TICv8 and spectroscopic information to determine stellar properties for the Giants Transiting Giants sample. Figure 2 shows an H-R diagram with evolutionary tracks downloaded from the MESA Isochrones \& Stellar Tracks (MIST; \cite{paxton2011,dotter2016,choi2016}). All Giants Transiting Giants systems were fit using \texttt{isoclassify} to highlight relative similarities between them. As all host stars have roughly the same mass and metallicity (M$_*$ $\approx$ 1.4 M$_\odot$, [Fe/H] $\approx$ 0.25 dex), we highlight this evolutionary track in red, and suggest that these systems may extend earlier analyses of ``retired A stars" to reveal evolutionary sequences for planetary systems orbiting post-main sequence, intermediate-mass stars \cite{johnson2007a,johnson2010}. We find that \hoststar\ is near the base of the red giant branch stage of evolution. 

In addition to the \texttt{isoclassify} determination of stellar parameters using \texttt{SpecMatch} derived effective temperature, stellar effective temperature, surface gravity, and metallicity were measured by combining multiwavelength photometry and parallax of \hoststar\ from the \emph{Gaia}, 2MASS, and WISE surveys. These values are then fit to MIST isochrones \cite{paxton2011,choi2016,dotter2016} to determine stellar mass, radius, and age. We find that our fundamental stellar parameter estimates from both approaches agree within uncertainties, and find that our statistical uncertainties on stellar parameters are smaller for our photometric analysis. We report the photometrically determined parameters in Table \ref{table:stellar} with model-dependent statistical uncertainties inflated to realistic values that better reflect the combination of statistical and systematic uncertainty \cite{tayar2022}, which we use for our subsequent system analysis. We list the derived stellar parameters and uncertainties in Table \ref{table:stellar}.

% In addition to the \texttt{isoclassify} determination of stellar parameters using \texttt{SpecMatch} derived effective temperature, log $g$ and metallicity, we also determine stellar parameters using multiwavelength photometry in the optical and infrared and a \emph{Gaia} parallax measurement, to more accurately characterize the stellar continuum and thereby calculate a more accurate effective temperature along with a self-consistent metallicity and log $g$ for this system. We then fit this multiwavelength photometry-determined effective temperature, log $g$ and metallicity to MIST isochrones to determine stellar mass, radius and age for this system. We use the parameters determined here for the subsequent mass loss analysis of this system, and list the derived stellar parameters and uncertainties in Table \ref{table:stellar}.

\subsection*{Planet Characterization} \label{sec:planetprops}

\subsubsection*{Model Fit}

We used the \exoplanet\ Python package to simultaneously fit a model to the photometry and radial velocity observations \cite{exoplanet:exoplanet}. The data input to our model were all Keck/HIRES radial velocity observations reported in this work and a \texttt{giants} light curve made from all sectors of \tess\ FFI photometry available from the first four years of the \tess\ Mission. Our model used stellar parameters derived from our multiwavelength photometric analysis following the procedure described above. We confirmed that this model was in strong agreement with an equivalent model fit using stellar parameters from \textsf{isoclassify} \cite{huber2017} with \emph{Gaia} parallax, and input effective temperature ($T_\mathrm{eff}$) and metallicity estimated using \texttt{SpecMatch} \cite{petigura2015} from spectral observations taken by the Keck-I/HIRES instrument. The photometric model input parameters used for our best-fit model can be found in Table \ref{table:stellar}. 

Our initial choices of planet period and depth were taken from the BLS search determined values produced during the transit search described in the \tess\ photometry section above. For limb darkening, we use the quadratic model prescribed by \cite{exoplanet:kipping13} to provide a two-parameter model with uninformative sampling. We parameterized eccentricity using the single planet eccentricity distribution of \cite{vaneylen2019}. 

We present our best fit models to the light curve and radial velocity data for \hoststar\ in Figure 1 and Table \ref{table:planet}. We present a more complete sample of the posteriors of our model in Figure \ref{fig:corner}.

 %Furthermore, we see additional deviations from standard planet transit models in at least the \planet system, which we discuss in more detail below.

Given that the transit signal detected in the `KSP SAP' and `SAP FLUX' light curves is less than twice the strength of the noise floor, and is not detected in the \texttt{eleanor} light curve, we only fit for planet parameters of \planet\ using the \texttt{giants} and SPOC light curves. We note that the best \texttt{exoplanet} model fit using the SPOC light curve gives a radius of 7.2 $\pm$ 0.8 R$_\oplus$, in good agreement with the value determined using the \texttt{giants} light curve. We find that the radial velocity signal for this planet is smaller than any other signal measured for a single transiting planet around an evolved star, but still constrains the planet mass to $>3\sigma$.

No significant out-of-transit variability can be resolved for \planet. Additional longer-baseline, higher-cadence data available for \planet\ from the extended \emph{TESS} Mission may provide evidence for variability in the stellar light curve which is currently not detectable. A long ($\sim$90 d) baseline of \emph{Kepler}-like precision photometry should easily allow detection of the asteroseismic signal of \hoststar \cite{grunblatt2016,grunblatt2017}. However, \hoststar\ is expected to be too faint to allow asteroseismic detection with \emph{TESS} \cite{huber2017}. We have produced a power spectrum of our \texttt{giants} light curve but cannot identify any asteroseismic power excess in the data.

\section*{Supplementary Text}

\subsection*{Planet Irradiation, Potential Mass Loss and Radius Inflation} \label{sec:inflation}

Previous to our \tess\ survey, only seven planets had been confirmed to be transiting evolved (T$_{\mathrm{eff}} <$ 6000 K, R $>$ 2 R$_\odot$) stars. These systems showed promise for solving mysteries of late-stage planet inflation and orbital evolution, but small numbers and undersampled parameter space has made determining population-wide characteristics difficult. With recent additional discoveries \cite{huber2019,rodriguez2021,khandelwal2022,saunders2022,grunblatt2022,wittenmyer2022,grunblatt2023}, \tess\ has now increased the number of confirmed planets in this population by over 100\% and has revealed a new regime of short-period hot Jupiters that have not yet inspiraled into their evolved host stars, constraining star-planet interaction rates \cite{villaver2014, hamer2019, attia2021}. In addition, our survey has revealed a number of other similar planet candidates which will be suitable for confirmation through ground-based follow-up observations in the near future. Measurements of the masses and eccentricities of these and similar systems will provide new constraints on planetary inflation, and atmospheric evolution that were not possible earlier \cite{grunblatt2017,chontos2019,komacek2020,thorngren2021,brinkman2023}. %We illustrate previously known and new planets, as well as new planet candidates orbiting evolved stars, in Figure \ref{fig:evolved_pop}.

% We illustrate planet radius as a function of incident flux in Figure \ref{fig:rad_v_flux}, highlighting the hot Neptune population. Color indicates planet mass, and we show the planets confirmed around evolved stars with stars (with the largest star corresponding to \planet).  

Relative to other known evolved systems, the incident flux on \planet\ is quite high. Assuming a direct correlation between planet radius and incident flux, it would be expected that this planet is among the most inflated Neptune-mass planets. As seen in Fig. \ref{fig:rad_v_flux}, the incident flux received by \planet\ is greater than that typically received by similarly-sized hot Neptunes---clustered to the left of \planet---by roughly a factor of 2. There are a few examples of similar planets that are more irradiated or more inflated that \planet, but this planet appears to fall deeper into the hot-Neptune desert as shown in incident flux and planet radius than any other planet currently known. Planets orbiting evolved stars appear to be somewhat overrepresented in the hot Neptune desert, featuring planets near both the upper and lower boundaries of the desert, but this may also be affected by observational and detection biases among this population.

Figures \ref{fig:rad_v_flux}, \ref{fig:mass_radius}, \ref{fig:density_massloss}, and \ref{fig:fractionalmassloss_mass} aim to place \planet\ in context of other, similar known planetary systems. Figure \ref{fig:rad_v_flux} shows planet radius as a function of incident flux for all planets with masses and radii reported with $<$50\% uncertainties on the NASA Exoplanet Archive. Color corresponds to planet mass, and the nominal threshold for inflation of 150 F$_\oplus$ \cite{demory2011} has been shown with a black dotted line. \planet receives a higher incident flux than any other planet with a radius between 0.5 and 0.8 Jupiter radii. Figure \ref{fig:mass_radius} shows radius versus mass for all planets with masses and radii reported with $<$50\% uncertainty on the NASA Exoplanet Archive. While not particularly extreme in either mass or radius, \planet\ is significantly more irradiated than most known Neptunes. In Figure \ref{fig:density_massloss}, we show the same population of planets in the planet density/atmospheric mass loss rate plane in the top panel, and the planet density/fractional atmospheric mass loss rate in the bottom panel. We find that in the density/mass loss rate plane, \planet\ is more similar to hot Jupiter systems than hot Neptunes, experiencing a very high current mass loss rate given its low mass and density. In the density/fractional mass loss plane, we have also labeled multiplanet systems with squares. \planet\ stands out as experiencing one of the highest fractional mass loss rates of planets with densities less than that of Jupiter. Interestingly, almost all planets expected to experience comparable amounts of mass loss are in multiplanet systems, where planet-planet interactions may have resulted in inflated planet radii \cite{millholland2019, millholland2020}. Figure \ref{fig:fractionalmassloss_mass} shows the same planet population in the fractional mass loss-planet mass plane. As in the previous Figure, we see that \planet\ is experiencing a higher fractional rate of mass loss than any other planet more massive than 10 Earth masses, with only rocky planets and super-puffs experiencing higher fractions of mass loss.

Given the large radii, relatively low densities and recent increases in irradiation due to host star evolution, these planetary systems can be interpreted as evidence for rapid re-inflation \cite{thorngren2021}, which could cause the large difference in planet radii among these particularly high equilibrium temperature planets. The vastly different planet radii could be due to different strengths and depths of various heat dissipation processes which could be at play for the planets studied here \cite{thorngren2018, komacek2020}. Differences in the bulk metallicity and the migration history of these planets will also increase scatter in planet radii as a function of mass and irradiation. The orbits currently observed for these planets may have been reached through tidal circularization induced by stellar evolution \cite{villaver2009,villaver2014}, which could also result in time-dependent internal heating and radius inflation due to the changing planetary orbit.

Atmospheric mass loss is also predicted to play a role in the sculpting of close-in planetary systems. High-energy EUV and X-ray photons can strip planetary atmospheres via photoevaporation \cite{murrayclay2009,schlaufman2021}. We use previously published relations \cite{king2021} between XUV and bolometric flux over the lifetimes of Sun-like stars to determine the expected cumulative atmospheric mass loss of all well-characterized planets with masses and radii measured to better than 50\% fractional uncertainty and an age reported on the NASA Exoplanet Archive. In Figure \ref{fig:xuvbol}, we illustrate the expected bolometric flux, XUV flux, and XUV to bolometric flux ratios over the course of a stellar evolutionary track for the Sun, as well as for \hoststar. We recreate the XUV to bolometric flux ratio law of Figure 1 in \cite{king2021} through a visual comparison, and use 100 Myr as the boundary between the saturated regime (where the flux ratio is constant, at early times) and the unsaturated regime (where the XUV flux declines steadily relative to the bolometric flux, at later times) \cite{johnstone2021}. We combine this with the reported semimajor axis to determine an XUV flux on any given planet, and then calculate an instantaneous mass loss rate using Equation 1 in the Main Text. We then integrate this mass loss rate over the expected age of the system to determine a cumulative mass loss rate, which we then compare to the current planet mass to create Figure 3 in the Main Text.

Through this calculation, we find that \planet\ likely experienced the highest fractional mass loss rate of any well-characterized planet between 1.6 and 9.5 R$_\oplus$. For a planet of this mass and temperature, these mass loss rates are expected to result in total loss of the planet's gaseous envelope \cite{lopez2014}. Furthermore, we find that another evolved system may also have experienced comparably high (or even higher) amounts of cumulative mass loss. KELT-11 b, a planet with a mass $<$0.2 M$_\mathrm{J}$ but a radius of 1.35 R$_\mathrm{J}$, does not have an age listed in the NASA Exoplanet Archive. However, we can infer an accurate and precise age based on the mass and late evolutionary stage of its host star following the same procedure as used for \hoststar. We find an age of 3.2 $\pm$ 0.1 Gyr, and use this age estimate to predict that KELT-11 b may also have lost over 65\% of its atmospheric mass, potentially the highest rate of mass loss for any gaseous planet. The fact that both of these cumulative mass loss outliers orbit evolved stars indicates that the evolutionary state of the systems may be related to their high cumulative expected mass loss. This suggests that late-stage evolutionary processes, such as rapid re-inflation, may have occurred, resulting in particularly high estimates of atmospheric mass loss for evolved systems due to the late-stage increase in planet radius. 

In order to ensure that photoevaporation is the dominant mass loss process occurring in this system, we also calculate the Roche lobe size for all planets in this sample, and compare the Roche lobe radius to the planet radius in Figure \ref{fig:rroche}. We find that very few well-characterized planets have radii that reach more than half their Roche lobe, and no planets that fill more than 60\% of their Roche lobe in our analysis. \planet\ appears to fill only 20\% of its Roche lobe, and thus is unlikely to be experiencing Roche lobe overflow. This is in line with \cite{jackson2017} and \cite{koskinen2022} indicating that Roche lobe overflow tends only to be important for hot Neptunes with $\lesssim$2 day orbital periods.

%Jeans escape or other atmospheric mass loss processes that similarly require Roche lobe filling or overflow.

\subsection*{Potential follow-up Observations}

\subsubsection*{Transit Spectroscopy}

Little is understood about the evolution of planetary atmospheres over time, and few observations of planetary atmospheres have been conducted for planets orbiting evolved stars \cite{kilpatrick2018,colon2020,mounzer2022}. Due to their particularly short orbits, the system introduced here is more well-suited for atmospheric characterization than other planets orbiting evolved stars.

Future observations with facilities capable of higher-precision photometry will confirm the true transit depth of this system, more precisely constraining the radius of \planet. Observations of the planet transit using a ground-based telescope with diffuser-assisted photometry \cite{stefansson2017} may also provide the requisite signal-to-noise ratio to obtain a more precise transit depth of \planet. Given the high scatter and low signal-to-noise ratio of this transit, planet parameters could also be significantly improved with transit observations taken by JWST \cite{gardner2006}. The higher photometric precision of JWST relative to TESS will allow better constraints on the planet radius, as well as transit shape and duration, which will help to constrain planetary orbit properties. Furthermore, the wavelength sensitivity of JWST may allow transmission spectroscopy. We find that \planet\ has a transmission spectroscopy metric (TSM; \cite{kempton2018}) value of $\sim$25, making it a challenging but accessible target for JWST transit spectroscopy observation. Preliminary predictions and comparisons to similar targets suggest that molecular signatures, clouds and/or hazes should be detectable in the planet's atmosphere \cite{mounzer2022, rustamkulov2022}. Recent observations have suggested a distinction in the atmospheric profiles of hot and ultra-hot Jupiters \cite{baxter2021}, which may also be seen between this planet and slightly cooler hot Neptunes. Additionally, atomic and molecular abundances measured from transmission spectroscopy may also probe where the planet originally formed and at what point in its lifetime it moved to its current orbit \cite{oberg2011,dawson2018}.

\subsubsection*{Radial Velocity}

Additional radial velocity observations of this system will also test dynamical evolution models. \cite{grunblatt2018} showed that giant planets orbiting giant stars at periods $<$30 days have average eccentricities $e>0.1$. However, at the shortest orbital periods ($<$5 days), even planets around evolved stars appear to have largely circular orbits. Constraints on orbital eccentricities will constrain both planet engulfment and stellar structure models \cite{weinberg2017, sun2018, soaresfurtado2021, grunblatt2023}. We find no evidence for significant eccentricity in the orbit of \planet\ based on our current set of observations. Using our best-fit \texttt{exoplanet} model of this system, we find that \planet\ has an eccentricity significantly smaller than $0.4$. A more finely sampled set of radial velocity measurements for this target with next generation instrumentation, such as Keck/KPF \cite{gibson2016} or WIYN/NEID \cite{schwab2016}, will place tighter constraints on the planet's eccentricity, and thus inform models of tidal evolution for the lowest mass ratio planetary systems, and determine whether this planet has completed any potential high-eccentricity migration \cite{villaver2014}. High-precision photometry detecting the secondary eclipse of the planet could provide a much more precise constraint on eccentricity.

%This relatively low RV jitter can be interpreted in different ways. This star may simply be a statistical outlier with relatively low RV jitter star for its evolutionary state. 

%This relatively low RV jitter can be interpreted in different ways. This star may simply be a statistical outlier with relatively low RV jitter star for its evolutionary state. 

The radial velocity jitter of \hoststar\ is measured to be 4.20 $\pm$ 0.90 m s$^{-1}$, in agreement with the 5 m s$^{-1}$ jitter typically expected for stars at this evolutionary state \cite{tayar2019}. Low radial velocity jitter may be related to lower stellar activity, which is likely also correlated with lower XUV flux for a star \cite{chadney2015}. Further radial velocity follow-up of this system with next generation extreme precision radial velocity instruments such as NEID and KPF may help to distinguish between stellar jitter and planetary signal in this system, improving our ability to characterize this planet as well as its host star.

\subsection*{Is \planet\ experiencing runaway inspiral?}\label{sec:orbdec}

The expected tidal interaction between hot gas giant planets and evolved host stars is expected to result in rapid orbital decay and eventual engulfment of the planet. However, orbital decay has only been measured in two systems to date, WASP-12 b and Kepler-1658 b \cite{yee2020,patra2020,turner2021,vissapragada2022}, both of which were less evolved than the system studied here. By constraining the rate of orbital decay in evolved systems, we can better characterize the strength of star-planet tidal interactions and their dependence on star and planet properties.

Figure \ref{fig:orbdec_pop} illustrates the population of known planets, highlighting those planets which are most likely to be experiencing strong orbital decay, as well as decay rates predicted using the prescription from \cite{goldreich1966}, assuming a modified stellar tidal quality factor $Q'_*$ = 10$^6$. The planets with the smallest relative orbital separations and highest masses relative to their stars decay most quickly, and can be found in the upper left hand corner of this plot. We have illustrated the new planets found by this survey as squares on this plot. These planets are among the best candidates for detecting orbital decay. In particular, \planet\ is predicted to have the most rapidly decaying orbit for a known planetary system with a mass ratio below 10$^{-4}$ to date. In addition, \planet\ also orbits a relatively cool star, which is expected to increase the speed of its orbital decay due to more rapid tidal dissipation in \hoststar's thick outer convective envelope \cite{patra2020}. However, the relatively noisy \emph{TESS} coverage for this system and relatively slow predicted orbital decay rate of this planet make it unlikely that orbital decay will be detectable for this system in the next decade.

% In Figure \ref{fig:planettides}, we illustrate the planet semimajor axis $a$ as a function of surface gravity in log solar units. The dotted line illustrates rough limits for the critical semimajor axis value, below which runaway inspiral is inevitable, as defined by \cite{sun2018}. As star-planet tidal interactions are more strongly sensitive to the ratio of semimajor axis to stellar radius than any other physical property \cite{1977A&A....57..383Z,hut1981, sun2018}, this diagram helps to highlight the most dynamically active systems, near the bottom of the plot. \planet\ is among a rapidly growing population of planets on the closest orbits ever found around stars $>$3 R$_\odot$. However, tidal interactions are also correlated with the planet mass, and thus the low mass of \planet\ may limit the strength of tides on this planet. 

To predict the orbital stability of this system at the current time, we use the formulation of \cite{sun2018} to determine the critical semimajor axis, $a_\mathrm{crit}$, at which the dynamical tides are expected to overcome equilibrium tides and begin the process of runaway inspiral begins. We compare $a_\mathrm{crit}$ to the current semimajor axis $a$ of the system, which we have determined to be 0.0622 $\pm$ 0.0049 au. Following the analytical approximations for $a_\mathrm{crit}$ assuming various rates of tidal dissipation which allow for different eddy sizes and turnover times within the convective envelope of the star, we determine a maximum $a_\mathrm{crit}$ = 0.055 au for this system assuming a non-reduced kinematic viscosity for this star \cite{zahn1977}. Thus, this planet is not predicted to be experiencing runaway inspiral and orbital decay at this time, and suggests that the host star must grow significantly in radius before this process begins.

% To predict the orbital stability of this system at the current time, we use the binary population synthesis package COSMIC \cite{breivik2020}. Using COSMIC allows us to predict the evolution of the host star and the orbital properties of the planet system as a function of time. In particular, we are interested to see if the orbital period of \planet\ is stable or actively decaying, based on the properties of this system. We note that COSMIC does not account for the effect of dynamical tides.

% We illustrate the results of this system evolution in Figure \ref{fig:orbevol}. We find that the orbit of \planet\ is stable until the host star reaches a radius of $\sim$4 R$_\odot$, larger than the 3.1 R$_\odot$ radius found for \hoststar\ at the current time. This implies that the orbit of \planet\ is currently stable and not decaying, but the planet's orbit is expected to begin decaying in $\approx$100 Myr from now, in agreement with the predictions of Figure \ref{fig:orbdec_pop}.

\clearpage

%Supp. Figures

\clearpage

\clearpage

\clearpage

\clearpage

\begin{figure*}
    \centering
    \includegraphics[width=\textwidth]{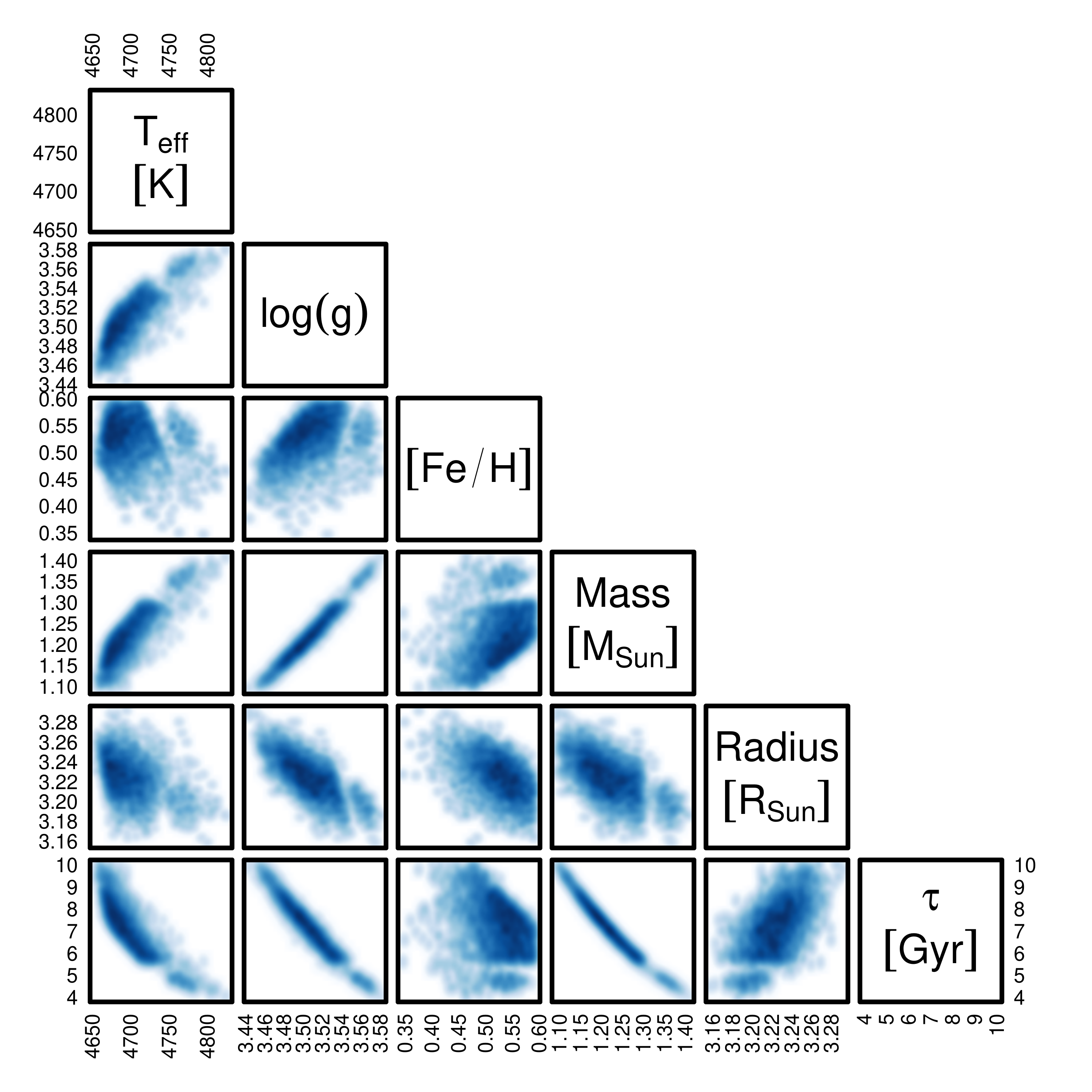}
    \caption{Corner plot illustrating the convergence of the stellar parameters determined for \hoststar\ by fitting the \emph{Gaia} DR2 $G$,$B_p$ and $R_p$, 2MASS J, H, and K, and WISE W1, W2, and W3 photometry as well as \emph{Gaia} parallax to MIST isochrones. Stellar mass, radius and age determined here are independently confirmed via a spectroscopic approach and are shown to agree within uncertainties.}
    \label{fig:mass_radius}
\end{figure*}

\clearpage

\begin{figure*}
    \centering
    \includegraphics[width=\textwidth]{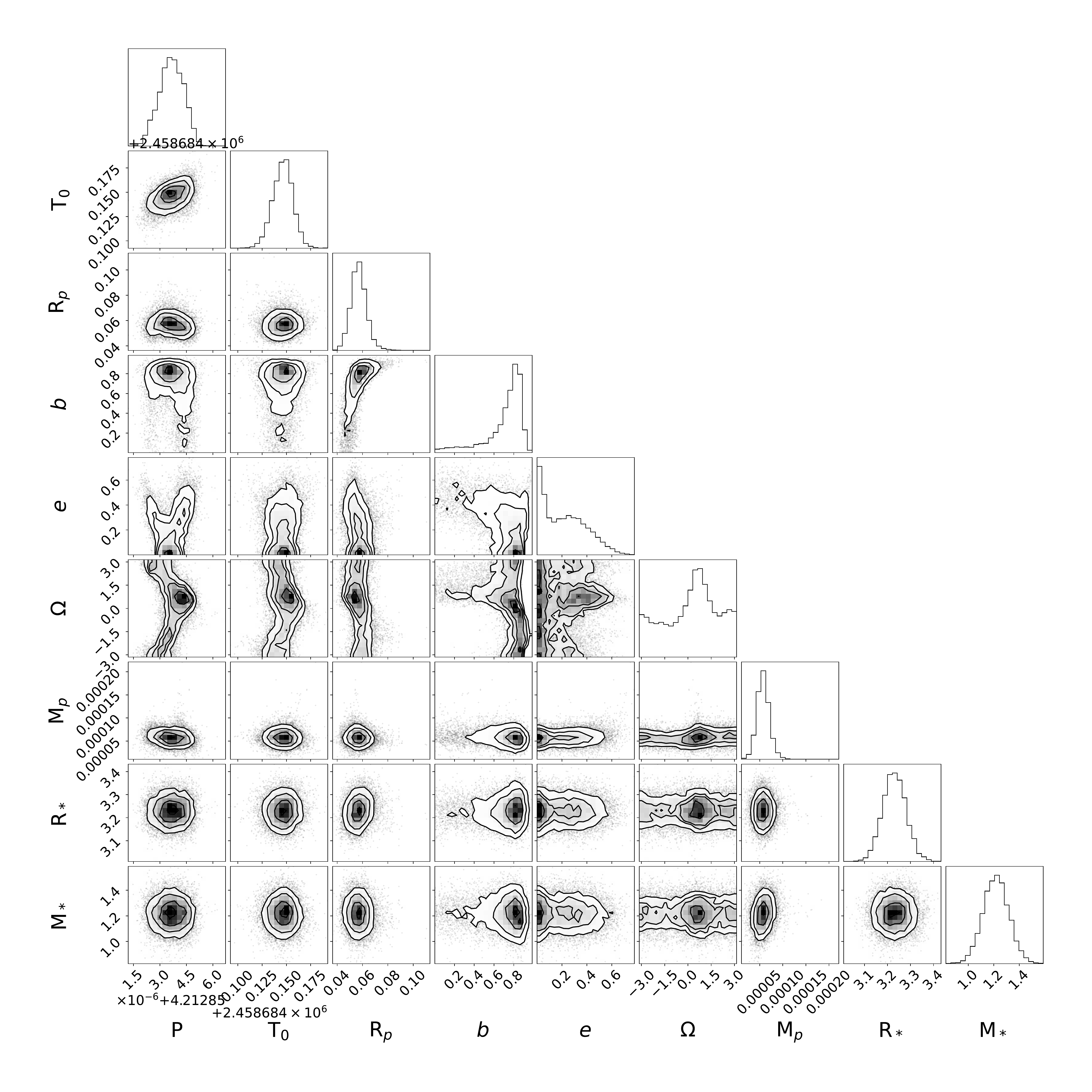}
    \caption{Corner plot illustrating the posteriors of the joint transit and RV model produced with \texttt{exoplanet}. The \texttt{giants} light curve has been used for this analysis. A total of 10 000 steps have been taken in order to produce these posterior chains. Planet mass and radius, and orbital period and ephemeris have clearly converged to unimodal distributions.}
    \label{fig:corner}
\end{figure*}

\clearpage

\begin{figure*}
    \centering
    \includegraphics[width=.85\textwidth]{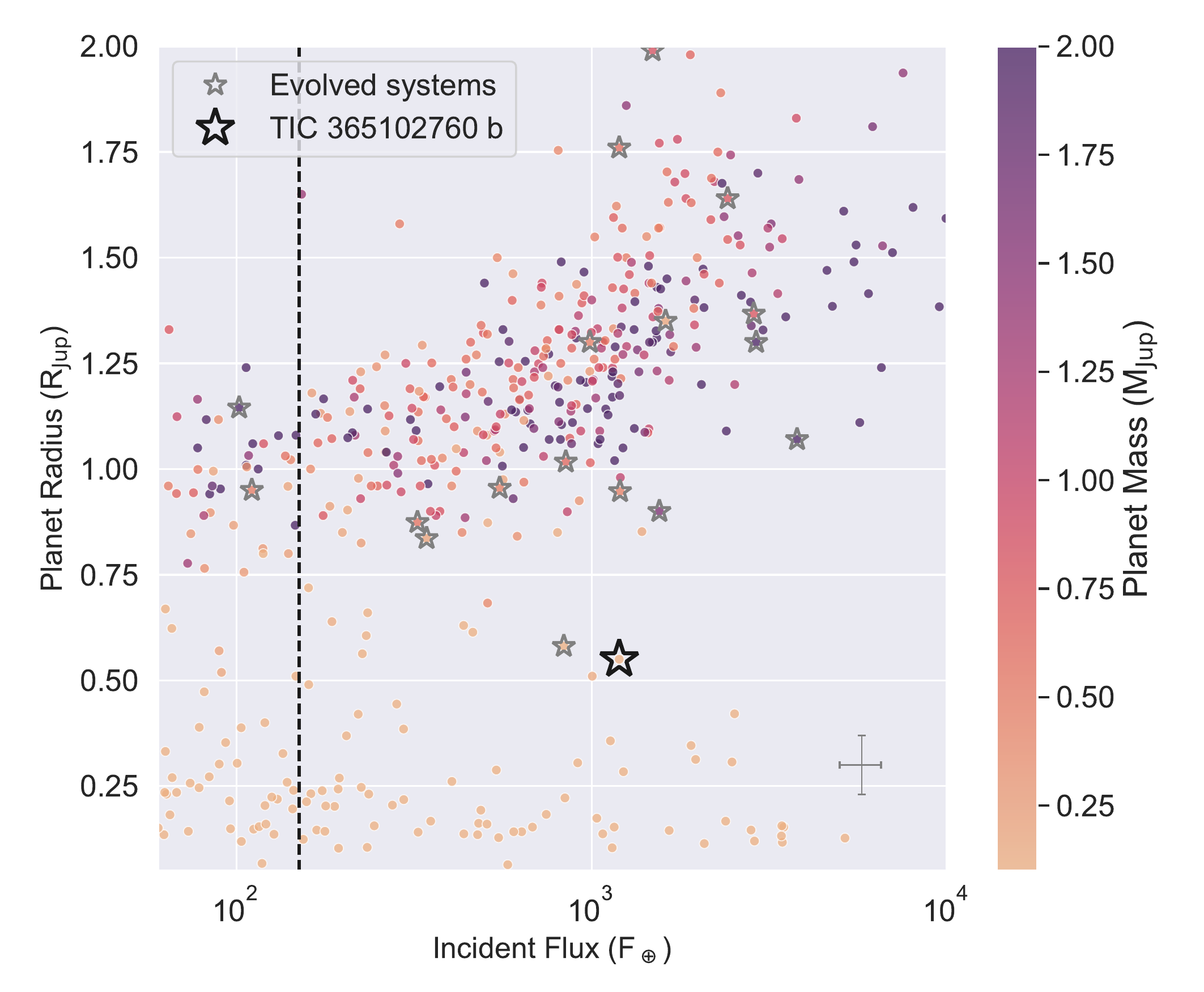}
    \caption{Planet radius vs. incident flux for all known planets with a well-characterized mass and radius, as listed on the NASA Exoplanet Archive on 31-Jan-2023. The black line corresponds to the observed 150 F$_\oplus$ threshold for planet inflation \cite{demory2011}. Color corresponds to planet mass. Typical uncertainties for the data are shown by the gray point in the lower right hand corner of the plot. The other previously known evolved, low-mass planet (R$_*$ $>$ 2 R$_\odot$, T$_\mathrm{eff}$ $<$ 6000 K) systems have been highlighted with gray stars on this plot, with \planet\ featuring a larger, black star. \planet\ is the most irradiated planet with a radius between 0.5 and 0.75 Jupiter radii, making it a valuable new benchmark for constraining atmospheric mass loss from stellar irradiation in Neptune-like planets. }
    \label{fig:rad_v_flux}

\end{figure*}

\clearpage

\begin{figure*}
    \centering
    \includegraphics[width=\textwidth]{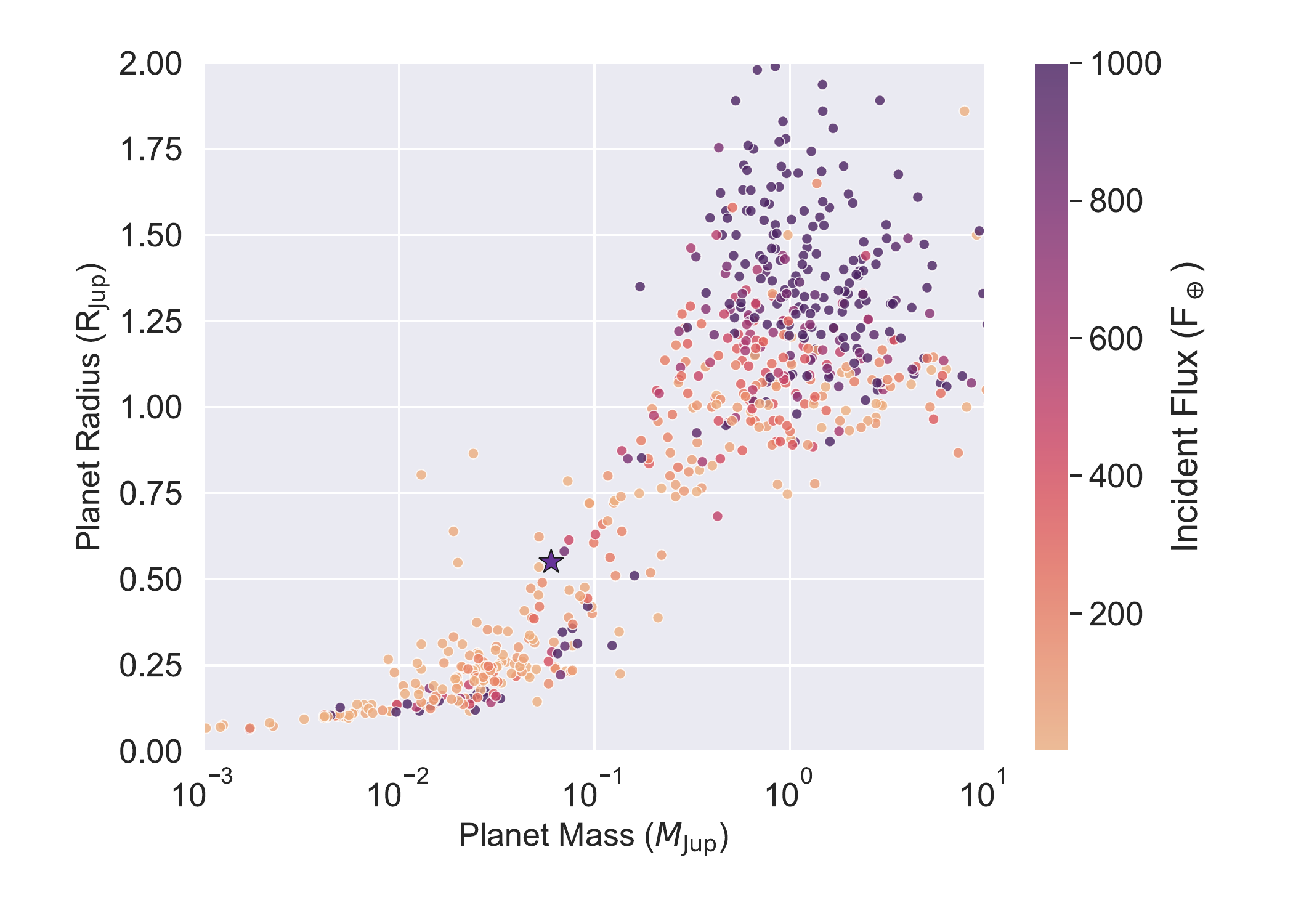}
    \caption{Planet mass versus planet radius for well-characterized confirmed planets, with color corresponding to incident flux. \planet\ is denoted by a star, and its high incident flux stands out among similarly-sized planets.}
    \label{fig:mass_radius}
\end{figure*}

\clearpage

% \begin{figure*}[ht!]
%     \centering
%     \includegraphics[width=\textwidth]{radflux_sns_103122.png}
%     \caption{Planet radius vs. incident flux for all known planets with a well-characterized mass and radius, as listed on the NASA Exoplanet Archive on 10-13-2022. The black line corresponds to the observed threshold for planet inflation \citep[150 F$_\oplus$;][]{demory2011}. Color corresponds to planet mass. Typical errors for the data are shown by the gray point in the lower right hand corner of the plot. The only other previously known evolved, low-mass planet (R$_*$ $>$ 2 R$_\odot$, T$_\mathrm{eff}$ $<$ 6000 K) systems have been highlighted with squares, whereas the new system confirmed by the \emph{TESS} GTG program, have been highlighted with stars on this plot. \planet appears to be among the most irradiated and most inflated planets (and the most irradiated evolved planet) on this figure, making it a valuable new benchmark for constraining atmospheric loss models.}
%     \label{fig:rad_v_flux}
% \end{figure*}

% \clearpage

\begin{figure*}
    \centering
    \includegraphics[width=.7\textwidth]{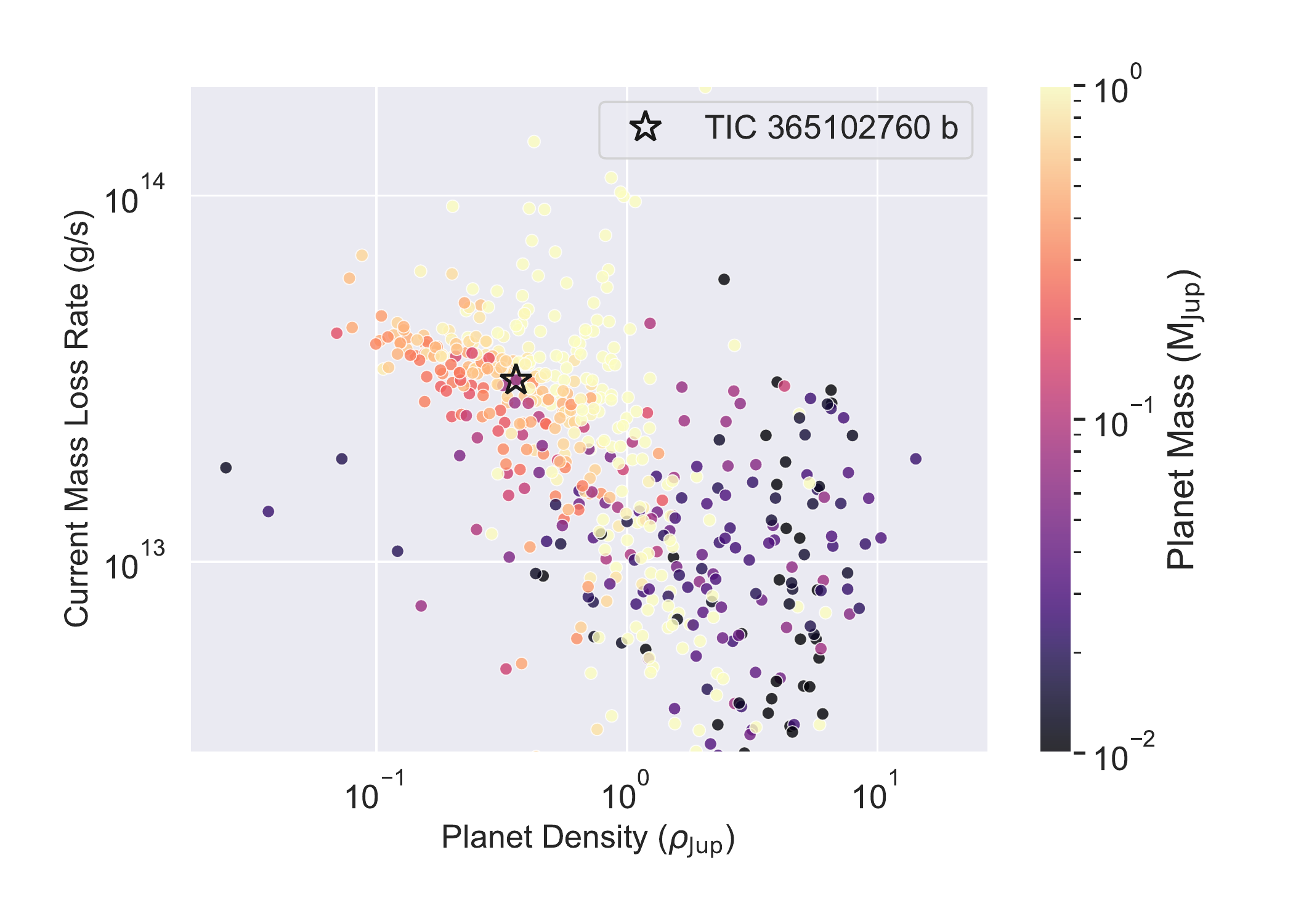}
    \includegraphics[width=.75\textwidth]{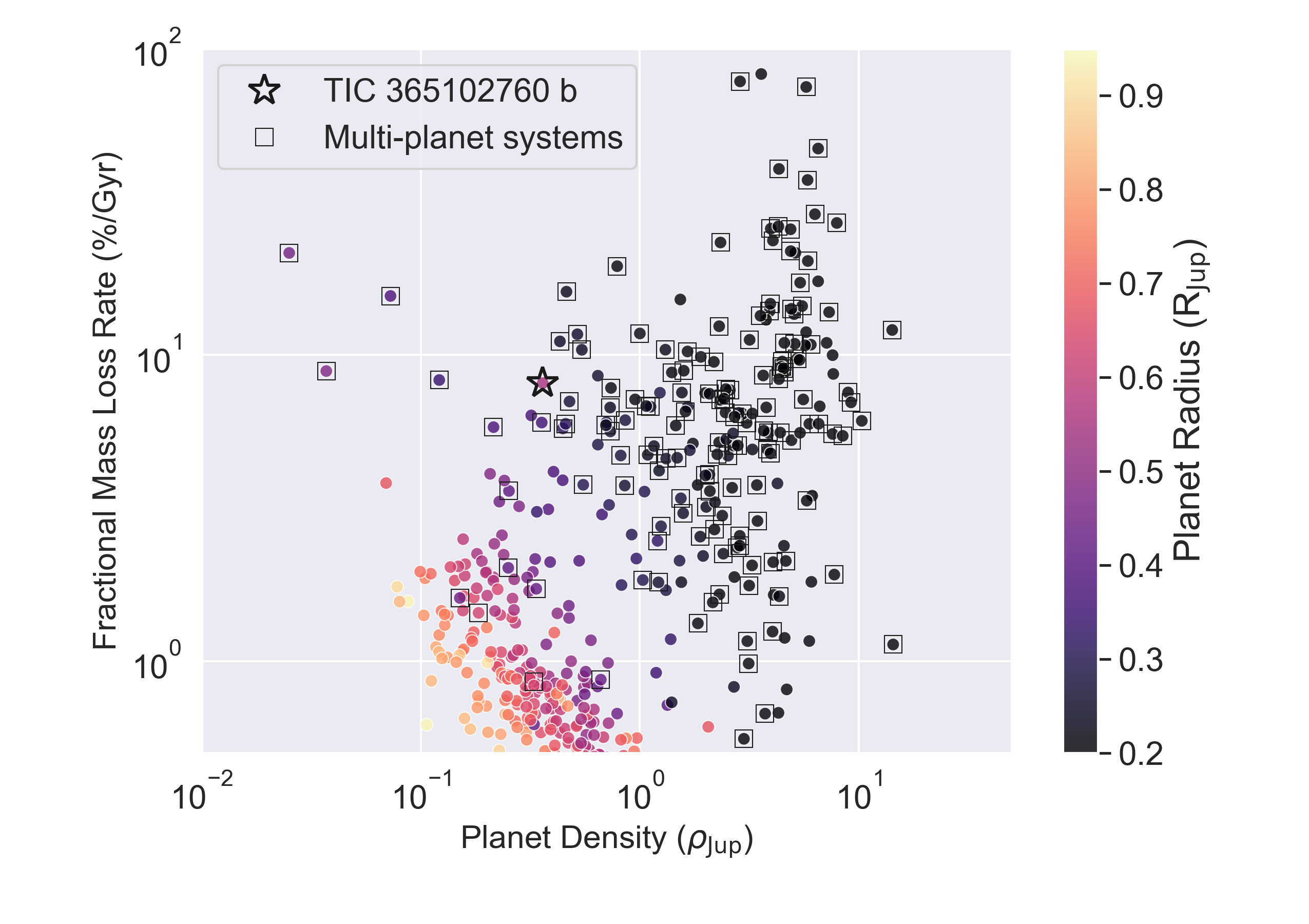}
    \caption{ {\it Top:} Atmospheric mass loss rate in grams/s versus planet density for all well-characterized planets, zoomed in to those experiencing the highest rates of mass loss. Color indicates the planet mass. \planet\ is denoted by a star, and stands out as relatively massive for its density and current mass loss rate. Planets with similar densities and mass loss rates tend to be much more massive than \planet. {\it Bottom:} Fractional mass loss rate in \%/Gyr versus planet density, where color represents planet radius. Planets in multiplanet systems have been designated by squares. \planet\ has the highest fractional mass loss rate of any planet larger than Neptune that is not in a multiplanet system (where orbital resonances can inflate planet radii).}
    \label{fig:density_massloss}
\end{figure*}

% \clearpage

% \begin{figure*}
%     \centering
%     \includegraphics[width=\textwidth]{fractionalmasslossrate_density_102822.png}
%     \caption{Orbital period versus planet radius for confirmed planets and new candidates transiting evolved (R$_*$ $>$ 2 R$_\odot$, T$_\mathrm{eff}$ $<$ 6000 K) stars. Planets known around evolved stars before the launch of \tess\ are shown in green. Those confirmed by \tess\ are shown as the largest symbols. Additional community-flagged planet candidates found by \tess\ are shown as small red stars, and \tess\ Objects  of Interest (TOIs) are shown in gray. \planet\ is shown by the largest star near the center of the Figure, and is the smallest planet found so far around an evolved star.}
%     \label{fig:tic3651_evoltracks}
% \end{figure*}

\clearpage

\begin{figure*}
    \centering
    \includegraphics[width=\textwidth]{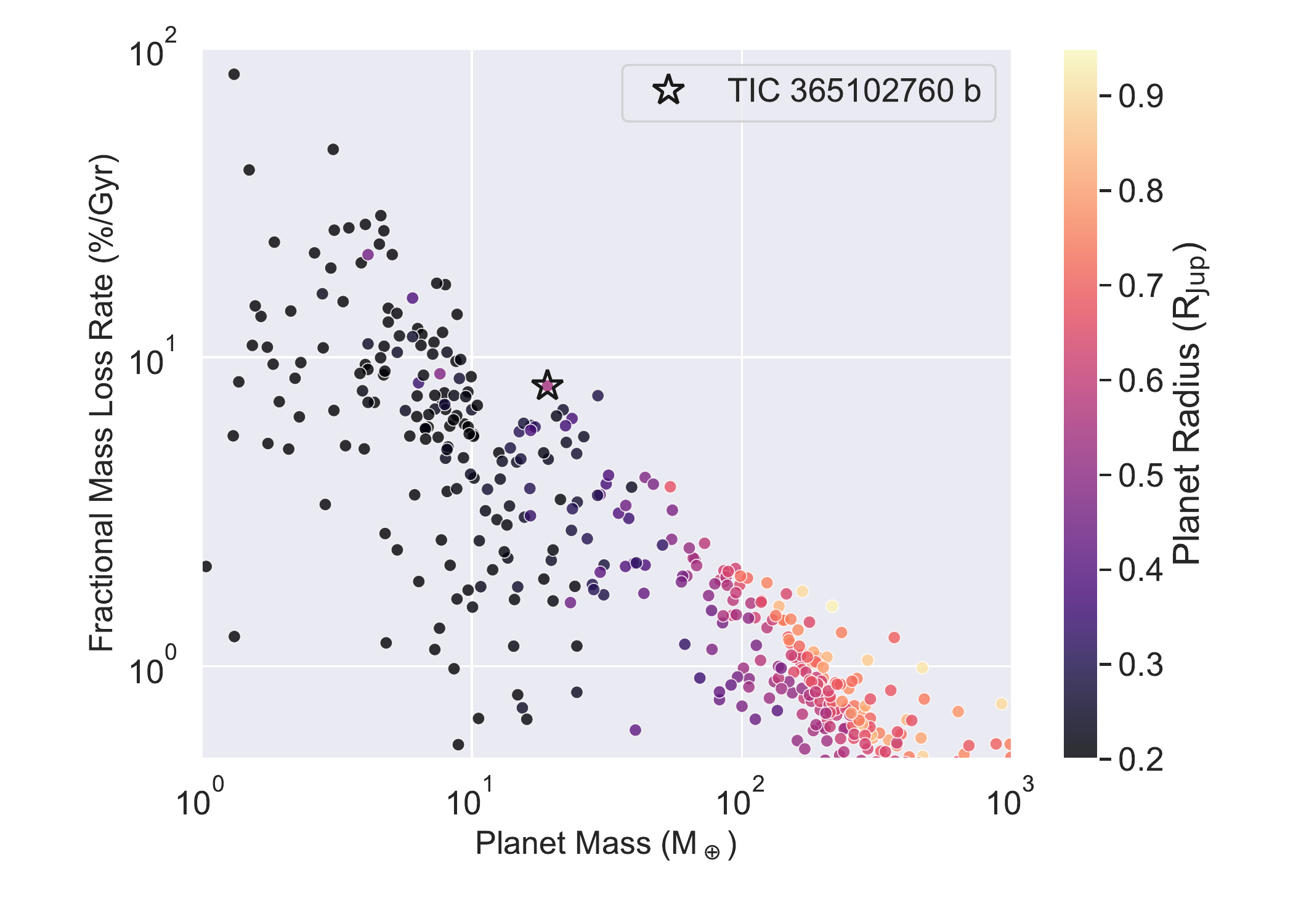}
    \caption{Fractional mass loss rate in \%/Gyr versus planet mass, where color represents planet radius. \planet\ has the highest fractional mass loss rate of any planet with a mass larger than 10 Earth masses, with only rocky planets and super-puffs experiencing higher fractional atmospheric mass loss rates.}
    \label{fig:fractionalmassloss_mass}
\end{figure*}

\clearpage

\begin{figure*}
    \centering
    \includegraphics[width=\textwidth]{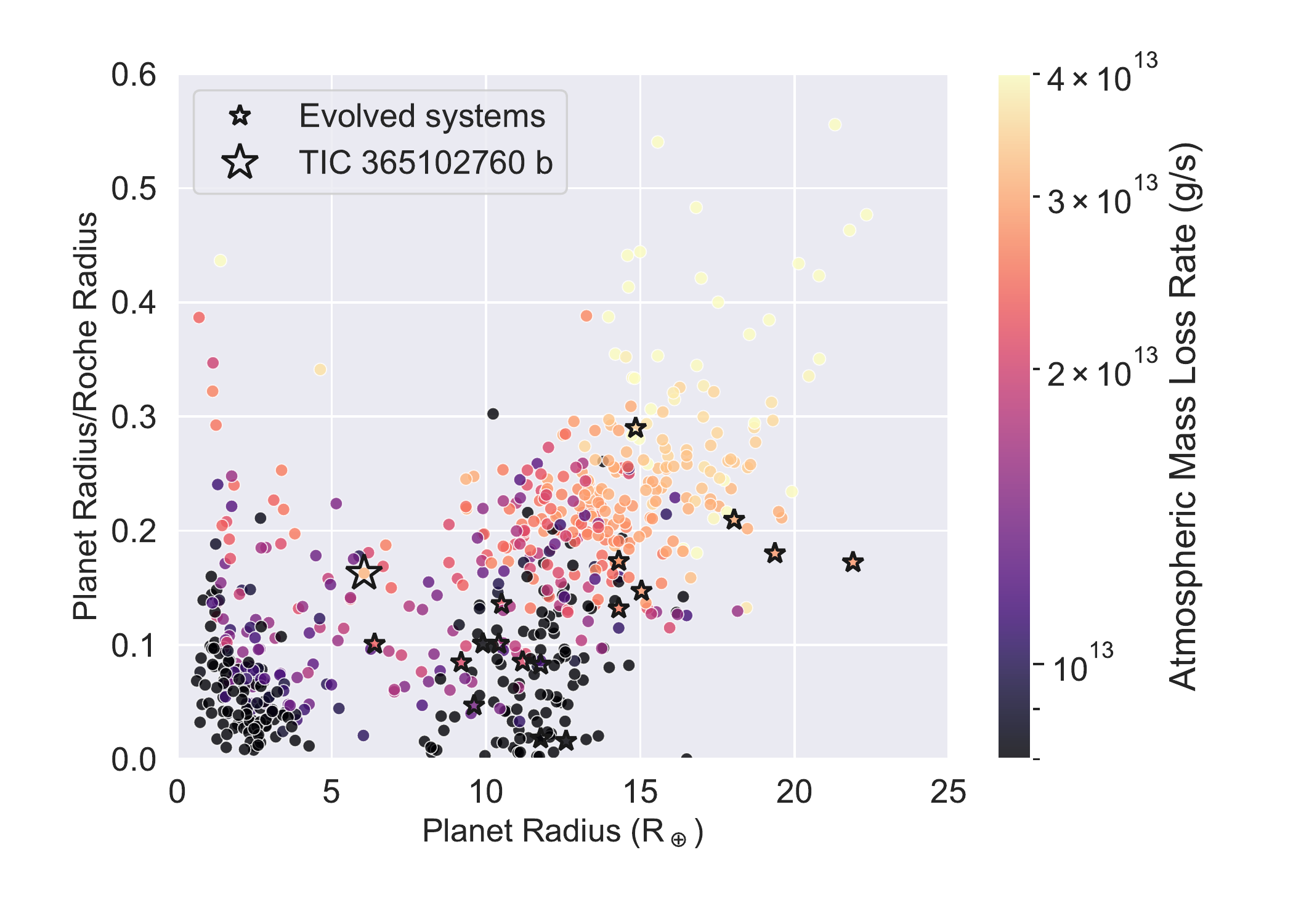}
    \caption{Ratio of planet radius to planetary Roche lobe as a function of planet radius. Evolved stars have been marked with star symbols, with \planet\ featuring the largest star. Color indicates the calculated rate of atmospheric mass loss. \planet\ is filling approximately 20\% of its Roche lobe, indicating it is unlikely to be experiencing any mass loss due to Roche lobe overflow.}
    \label{fig:rroche}
\end{figure*}

\clearpage

\begin{figure*}
    \centering
    \includegraphics[width=\textwidth]{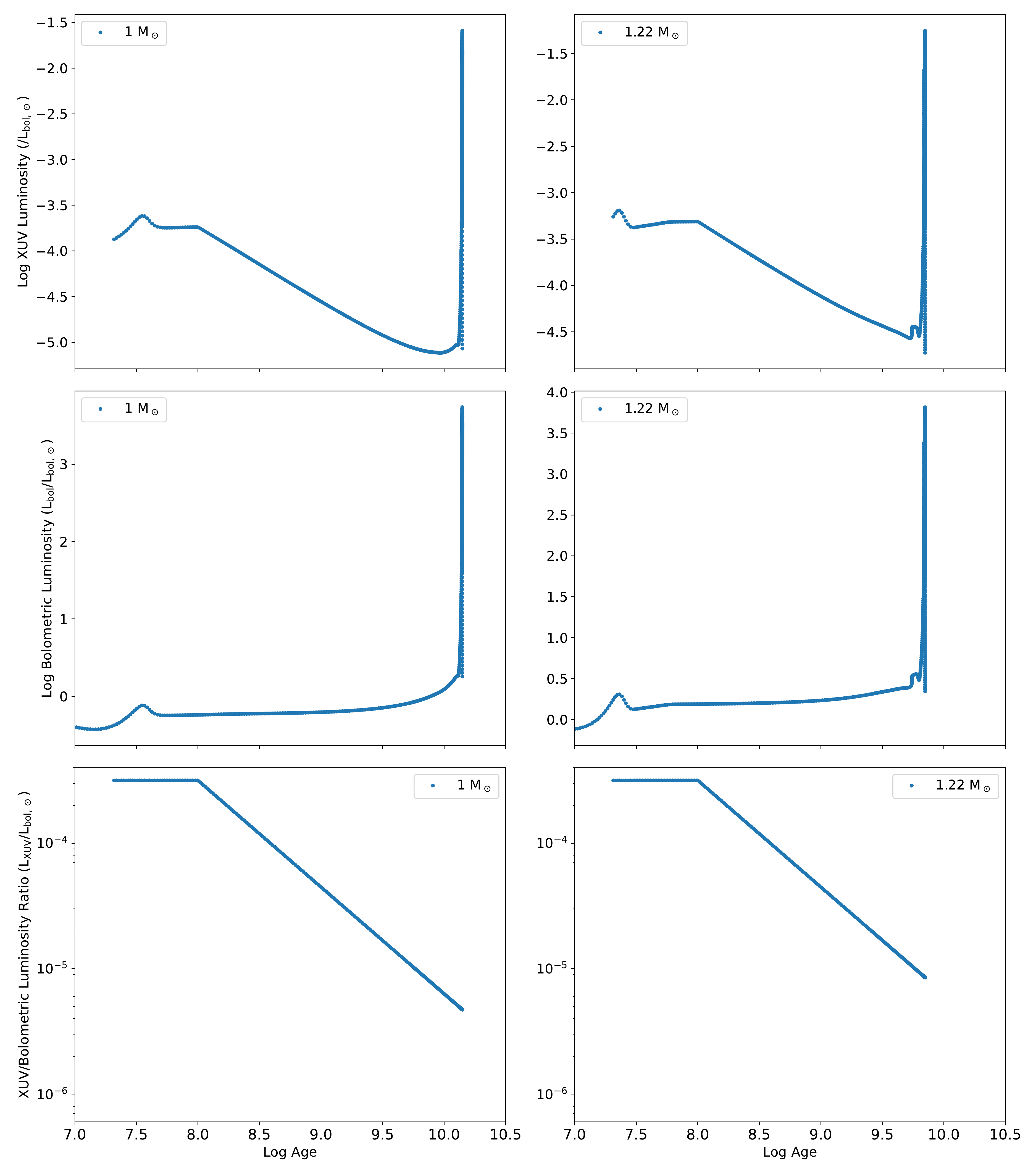}
    \caption{{\it Top row:} XUV luminosity versus the logarithm of the stellar age in years for 1 M$_\odot$ (left) and 1.22 M$_\odot$ (right) stellar models. The luminosity is defined with respect to the bolometric luminosity of the Sun. {\it Middle row:} Bolometric luminosity for 1 M$_\odot$ (left) and 1.22 M$_\odot$ (right) stellar models, also defined with respect to the bolometric luminosity of the Sun. {\it Bottom row:} Ratio of XUV flux to bolometric flux as a function of age for 1 M$_\odot$ (left) and 1.22 M$_\odot$ (right) stellar models.}
    \label{fig:xuvbol}
\end{figure*}

\clearpage

\begin{figure*}
    \centering
    \includegraphics[width=\textwidth]{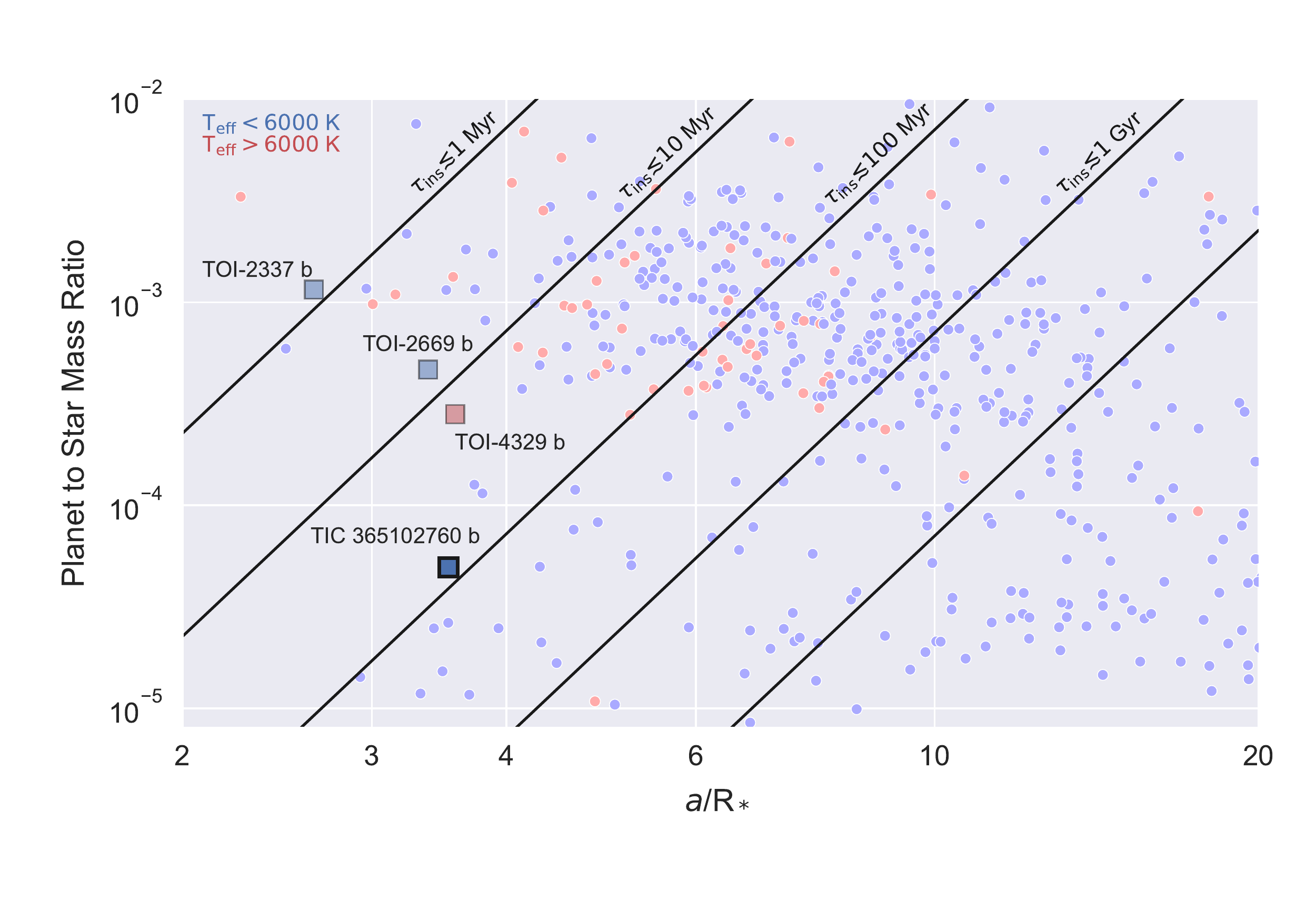}
    \caption{Semimajor axis divided by stellar radius, versus planet to star mass ratio for confirmed planets. Orbital decay timescales are determined assuming a tidal  decrease toward the upper left of this plot, where black diagonals correspond to theorized rates of orbital decay, where the leftmost line corresponds to a decay timescale of 10$^6$ years, and each following line increases by a factor of 10. Blue points have stellar effective temperatures $<$6000 K as reported by the NASA Exoplanet Archive, while red points represent planets around hotter stars. The planets confirmed by our \emph{TESS} GTG II paper \cite{grunblatt2022} are shown as squares with black outlines, and are populating relatively sparse regions of parameter space on this plot that correspond to rapid orbital decay. \planet\ may be experiencing the fastest rate of orbital decay of any planetary system known with a mass ratio below 10$^{-4}$ or a planet mass $<$25 M$_\oplus$.}
    \label{fig:orbdec_pop}
\end{figure*}

% \clearpage

% \begin{figure*}[ht!]
%     \centering
%     \includegraphics[width=\textwidth]{a_logg_3651.png}
%     \caption{
%     Semimajor axis vs. log($g$) for the confirmed planet population, highlighting planets from our \emph{TESS} Giants Transiting Giants programs as red stars. \planet is the fourth system found by \emph{TESS} at or beyond the critical semimajor axis where runaway inspiral begins.}
%     \label{fig:planettides}
% \end{figure*}

% \clearpage

% \begin{figure*}
%     \centering
%     \includegraphics[width=.49\textwidth]{cosmic_rstar.png}  \includegraphics[width=.49\textwidth]{cosmic_sep.png}
%     \caption{Stellar radius (left) and orbital separation (right), as a function of time for the \hoststar\ system. Using COSMIC to simulate this planetary system as a evolving binary, we are able to determine the time and expected stellar radius at which \planet\ is expected to experience runaway inspiral. We find that the orbital period of this system is constant until the star reaches a radius of $\gtrsim$4.5 R$_\odot$, implying that the planet's orbit is not yet decaying at an increasing rate. Stellar winds do not appear to strongly affect this rate or timing of inspiral. Changes in metallicity relative to the Sun may cause inspiral to be delayed by tens of millions of years, but the relative amount of time to inspiral from the current state of the system is largely unchanged.}
%     \label{fig:orbevol}
% \end{figure*}

\clearpage

%Supp. Tables

\begin{table}[]
    \centering
    \begin{tabular}{c c}
        \hline
        Time (JD - 2457000) & Relative RV (m/s)  \\
        \hline
        2379.061 & 0.3 $\pm$ 1.5 \\
        2386.098 & -7.7 $\pm$ 1.4 \\
        2396.065 & 1.1 $\pm$ 1.5 \\
        2413.082 & 8.0 $\pm$ 1.7 \\
        2421.011 & -2.8 $\pm$ 1.9 \\
        2423.062 & -2.8 $\pm$ 1.4 \\
        2435.909 & -2.0 $\pm$ 1.3 \\
        2442.078 & 5.1 $\pm$ 1.5 \\
        2446.082 & 1.4 $\pm$ 1.6 \\
        2449.040 & -0.7 $\pm$ 1.5 \\
        2503.884 & -14.9 $\pm$ 1.5 \\
        2509.902 & 10.8 $\pm$ 1.7 \\
        2513.860 & 8.0 $\pm$ 1.5 \\
        2545.776 & -6.4 $\pm$ 1.8 \\
        \hline
    \end{tabular}
    \caption{Radial velocities and uncertainties measured for \hoststar\ from Keck/HIRES. The RVs have been sorted in time.}
    \label{table:rvs}
\end{table}

\clearpage

\begin{table*}
\centering
    \begin{tabular}{l c c c}
        \hline
        \rule{0pt}{3ex}\textit{Target ID} & & & \\
        \rule{0pt}{3ex}        TIC & 365102760 &  &  \\
        2MASS & J20232153+5423395 &  & \\
        Gaia DR2 & 2185044477033336064 &  & \\
        \hline
        \rule{0pt}{3ex}\textit{Observables} & \\
        \rule{0pt}{3ex}RA(J2015.5) & 20:23:21.56 & & \\
        Dec(J2015.5) & 54:23:39.55 &  & \\
        B mag & 13.034 $\pm$ 0.526 &  & \\
        V mag & 12.154 $\pm$ 0.034 &  & \\
        \emph{Gaia} mag & 11.9772 $\pm$ 0.0002 &  & \\
        \tess\ mag & 11.33 $\pm$ 0.01 &  & \\
        2MASS J mag & 10.406 $\pm$ 0.026 &  & \\
        2MASS H mag & 9.894 $\pm$ 0.032 &  & \\
        2MASS K mag & 9.788 $\pm$ 0.020 &  & \\
        WISE W1 mag & 9.721 $\pm$ 0.023 &  & \\
        WISE W2 mag & 9.813 $\pm$ 0.020 &  & \\
        WISE W3 mag & 9.700 $\pm$ 0.034 &  & \\
        Proper Motion, right ascension $\mu_\mathrm{RA}$  & 10.8797 $\pm$ 0.0418 mas yr$^{-1}$ &  & \\
        Proper Motion, declination $\mu_\mathrm{dec}$  & 5.1928 $\pm$ 0.0422 mas yr$^{-1}$ &  & \\
        Radial Velocity & -1.59 $\pm$ 0.25 km s$^{-1}$ &  & \\
        Distance & 555.5 $\pm$ 7.6 pc &  & \\
        \hline
        \rule{0pt}{3ex}\textit{Inferred Characteristics} & \\
        
        Radius $R_\star$  & \starradius &  & \\
        Mass $M_\star$  & \starmass &  & \\
        $T_{\rm eff}$  $ $ & \teff & & \\
        $\log(g)$  & \logg & &  \\
        $ $[Fe/H]  $ $ & \feonh &  &  \\
        Age  $ $ & \age &  & \\
        Density $\rho_\star$  & \starrho & & \\
        \hline
   \end{tabular}
	 \caption{Stellar properties derived from a MIST isochrone fit to multiwavelength photometry from \tess\, \emph{Gaia}, 2MASS and WISE observations. All parameters are in good agreement with an independent determination of stellar parameters using \texttt{SpecMatch}-derived parameters of Keck-I/HIRES spectra with \texttt{isoclassify}.}
	 \label{table:stellar}
\end{table*}

\clearpage

\begin{table*}
\begin{center}
    \begin{tabular}{l c r}
        \hline 
        Parameter & Prior & Value \\
        \hline
        \rule{0pt}{3ex}\textit{Transit Fit Parameters} & & \\
        \rule{0pt}{3ex}Orbital period $P_{\text{orb}}$ [days] & $\log\mathcal{N}[4.2129, 0.001]$ &  \period \\
        Semimajor axis $a$ [AU] &  &  0.0622 $\pm$ 0.0049 \\
        Transit epoch $t_0$ [BJD - 2457000] & $\mathcal{N}[1684.177, 0.03]$ & \transittime \\
        Transit duration $T_\mathrm{dur}$ [hr] & & 5.52$_{-0.72}^{+1.68}$\\
        Planet-to-star radius ratio R$_p$/R$_*$ & & 0.0176 $\pm$ 0.0019\\
        Impact parameter $b$ & $P_\beta(e\in[0,1])^\text{(a)}$  & $0.704\pm0.199$\\
        Eccentricity $e$ & single-planet dist. from \cite{vaneylen2019} & $<$ 0.406 \\
        Argument of periastron $\Omega$ & $\mathcal{U}[-\pi, \pi]$& 0.216 $\pm$ 1.698 \\
        Limb-darkening coefficient $q_1$ & [0,2]$^\text{(b)}$ & $0.752 \pm 0.498$ \\
        Limb-darkening coefficient $q_2$ & [-1,1]$^\text{(b)}$ & $-0.056 \pm 0.424$ \\
        \hline
        \rule{0pt}{3ex}\textit{Radial Velocity Fit Parameters} & & \\
        \rule{0pt}{3ex}Semi-amplitude $K$ [m/s] & $\mathcal{U}[0, 50]$ & $7.4 \pm 1.6$\\
        RV jitter $\sigma_\mathrm{RV}$ [m/s] & $\mathcal{U}[0, 10]$ & $4.20 \pm 0.90$\\
        Offset $\gamma$ [m/s] & $\mathcal{U}[-100, 100]$ & 0.81 $\pm$ 2.54 \\
        \hline
        \rule{0pt}{3ex}\textit{Derived Physical Parameters} & & \\
        \rule{0pt}{3ex}Planet radius $R_p$ & $\mathcal{U}[0, 30]$ & \planetradius \\
        Planet mass $M_p$ & $\mathcal{U}[0, 300]$ & \planetmass \\
        Planet density $\rho_p$  & & 0.437$^{+0.229}_{-0.150}$ g cm$^{-3}$ \\ 
        $a/R_*$  & & 4.14 $\pm$ 0.35 \\ 
        \hline 
   \end{tabular}
	 \caption{Fit and derived parameters for \planet. Parameters without priors were inferred analytically through combinations of other parameters listed here. \textit{Note:} $^\text{(a)}$This parameterization is described by the Beta distribution in \cite{kipping2013b}. $^\text{(b)}$  Distributions follow correlated two-parameter quadratic limb-darkening law from \cite{kipping2013}.}
	 \label{table:planet}
\end{center}
\end{table*}

% \clearpage

% \bibliography{scibib}

\end{document}